%% file: hybridisation_Hoare_logic.tex
\documentclass[runningheads,a4paper]{llncs}

\usepackage[utf8]{inputenc}
\usepackage[T1]{fontenc}
\usepackage{amssymb}
\usepackage{stmaryrd}
\usepackage{mathtools}
\usepackage{graphicx}
\usepackage{amsmath}
\usepackage{xcolor}

\usepackage{prettyref}
\newrefformat{ssec}{Subsec.~\ref{#1}}
\newrefformat{sec}{Sec.~\ref{#1}}
\newrefformat{ap}{Appendix~\ref{#1}}
\newrefformat{def}{Def.~\ref{#1}}
\newrefformat{fig}{Fig.~\ref{#1}}
\newrefformat{ex}{Ex.~\ref{#1}}
\newrefformat{eq}{\eqref{#1}}
\newcommand{\pref}[1]{\prettyref{#1}}

\usepackage{url}
\urldef{\mailsa}\path|{alfred.hofmann, ursula.barth, ingrid.haas, frank.holzwarth,|
\urldef{\mailsb}\path|anna.kramer, leonie.kunz, christine.reiss, nicole.sator,|
\urldef{\mailsc}\path|erika.siebert-cole, peter.strasser, lncs}@springer.com|    
\newcommand{\keywords}[1]{\par\addvspace\baselineskip
\noindent\keywordname\enspace\ignorespaces#1}

\newcommand{\pii}[2]{{\pi_{#1}^{#2}}}
\newcommand{\etai}[2]{{\eta_{#1}^{#2}}}

\newcommand{\sN}{\mathbb{N}}
\newcommand{\N}{\sN}
\newcommand{\sR}{\mathbb{R}}
\newcommand{\R}{\sR}
\newcommand{\Nsegm}[2]{\llbracket #1, #2 \rrbracket}
\newcommand{\Rsegm}[2]{[#1, #2]}

\newcommand{\pred}{R^{-}}
\newcommand{\A}{\mathcal{A}}
\newcommand{\Junc}{\mathcal{J}}
\newcommand{\Rep}{\Junc}
\newcommand{\IW}{\mathsf{IW}}
\newcommand{\BW}{\IW}
\newcommand{\EW}{\mathsf{EW}}
\newcommand{\EB}{\EW}
\newcommand{\F}{\mathcal{F}}
\newcommand{\W}{\mathcal{W}}

\newcommand{\slide}{\mathsf{slide}}

\newcommand{\Sl}{\mathcal{S}}

\newcommand{\WP}{\mathsf{WP}}
\newcommand{\emptyseq}{\varepsilon}
\newcommand{\sgn}{\mathrm{sgn}}

\newcommand{\propinline}[2]{(#1, #2)}
\newcommand{\prop}[2]{\begin{Bmatrix}#1\\#2\end{Bmatrix}}
\newcommand{\patha}[3]{\begin{pmatrix}#1\\#2\\#3\end{pmatrix}}
\newcommand{\stackedconcat}[1]{%
\!\!\!\!\!\!\!\!\!\!
\begin{array}[t]{c}
  ;\\\\
  \uparrow \\
  #1\\
\end{array}\!\!\!\!\!\!\!\!\!\!}
\newcommand{\hoare}[3]{\{#1\}~#2~\{#3\}}

\newcommand{\resp}{resp.~}

\begin{document}

\mainmatter  

\title{A Hybrid Hoare Logic for Gene Network Models} 

\titlerunning{A Hybrid Hoare Logic for Gene Network Models
  }

%
%
\author{Jonathan Behaegel%
\and Jean-Paul Comet 
\and Maxime Folschette
\thanks{alphabetically ordered authors}
}
\authorrunning{Jonathan Behaegel, Jean-Paul Comet, and Maxime Folschette}

\institute{
University Nice Sophia Antipolis and 
I3S laboratory, UMR CNRS 7271\\
CS 40121, 06903 Sophia Antipolis CEDEX, France\\
behaegel@i3s.unice.fr, \{comet,maxime.folschette\}@unice.fr
}

%
%

\toctitle{Lecture Notes in Computer Science}
\tocauthor{Authors' Instructions}
\maketitle

\begin{abstract}
The main difficulty when modelling gene networks is the identification
of the parameters that govern their dynamics. It is particularly
difficult for models in which time is continuous: parameters have real
values which cannot be enumerated. The widespread idea is to infer new
constraints that reduce the range of possible values.  Here we present
a new work based on a particular class of Hybrid automata (inspired by
Thomas discrete models) where discrete parameters are replaced by
signed celerities. We propose a new approach involving Hoare
logic and weakest precondition calculus (a la
Dijkstra) that generates constraints on the parameter values. Indeed,
once proper specifications are extracted from biological traces with
duration information (found in the literature or biological
experiments), they play a role similar to imperative programs in the
classical Hoare logic. We illustrate our hybrid Hoare
logic on a small model controlling the
\emph{lacI} repressor of the lactose operon.

\end{abstract}

\input{intro}

\input{framework}

\input{hoare12}

\input{hoare3}

\input{example}

\input{conclusion}

\subsubsection*{Acknowledgements.} 
We are grateful to E. Cornillon and G. Bernot for fruitful discussions
about the hybrid formalism. This work is partially supported by the
French National Agency for Research (ANR-14-CE09-0011).

\bibliographystyle{plain}
\bibliography{hybridisation_Hoare_logic}

\appendix
\section*{Appendix}
\input{formulas}

\input{appendix}

\end{document}

%% file: intro.tex
\section{Introduction}
\label{sec:intro}



Regulatory networks are models based on gene regulation in the aim of
representing their functionning. They are “on/off” switching systems
of genes whose role is to control when each gene acts and when its
corresponding protein is synthesised.
When a gene switches on/off, its protein is synthesised or degraded
leading to a change of gene regulations. 
The main difficulty of such networks is the identification of parameters,
whatever the modelling framework (differential, qualitative or
stochastic models). These parameters govern the dynamics of the
system, and constraints should be identified at least to narrow
the set of suitable parameters. 
We need to accurately identify these parameters especially when time
plays an essential role like in the the coupling between the cell
cycle and the circadian clock.

Previous studies used René Thomas' discrete modelling
framework~\cite{TK2001-b} to determine the dynamics of such
systems~\cite{confCSBio2012}. This modelling is based on
discretisation of concentration spaces, so each threshold separates
concentration areas where the regulations are identical (the
regulations are supposed to be sharp sigmoid).

The hybrid framework we use in this paper describes concentration
levels in a qualitative way as in Thomas' discrete modelling, but
it also takes into account temporal information represented by
evolution speeds (called celerities).  The time spent inside each
discrete state can be approximated through biological data, thus
increasing the constraints on parameters and improving the exactness
of the model~\cite{chapterEvry2016-b}. Such a hybrid model of the cell
cycle was created~\cite{behaegel2016hybrid} and, using a suitable set
of parameters, we produced simulations in accordance with the
classical known behaviour of the cell cycle.

In this paper we use Hoare logic in order to establish constraints
on the celerities. Similar approaches were already developed for
Thomas' discrete approach~\cite{inviteCMSB2015,arXiv} and extensions
on more generic hybrid automata have been proposed
\cite{Martin2006,Ishii2013,Zou:2013:VSD:2555754.2555763}.
Hoare logic relies on so-called Hoare triples of the from
$\hoare{Pre}{p}{Post}$ (where $Pre$ and $Post$ are formulas and $p$ is a
path)~\cite{logique-Hoare-1969}
which mean that, starting from a
state satisfying the precondition $Pre$ and crossing the path $p$, the
trajectory reaches a point where the postcondition $Post$ is satisfied.
However, in this work, we mainly focus
on the computation of the weakest precondition $Pre$
for a given path $p$ and postcondition
$Post$~\cite{Dijkstra:1975:GCN:360933.360975}.

So, using
biological data, represented into a Hoare triple, we are able to
establish new constraints on parameters allowing the model to behave
as the observed traces. The description of the observed traces
essentially consists in depicting the correct order of the events
(i.e., the order of crossings of thresholds).  According to the
situation, we can also use other information, such as the
saturation or complete degradation of a protein which lead to
additional constraints.

We illustrate our approach on a genetic construct in \textit{Escherichia
  Coli} composed of the lactose operon
\textit{lacI} and some elements of the system \textit{Ntr}~\cite{atkinson2003development}.  The
\textit{glnA} promoter of system \textit{Ntr} is modified by adding the
operator site of the \textit{lacI} repressor (inhibition).  Similarly
the \textit{glnK} promoter of \textit{Ntr} is fused to the \textit{lacI}
gene (activation). These alterations lead to an oscillatory behaviour
in \textit{E.~Coli}. We deduce with our weakest precondition calculus
the constraints on celerities mandatory to observe
oscillation.

The paper is organised as follows. We first define the hybrid
modelling framework based on Thomas' discrete one
(\pref{sec:framework}). Then \pref{sec:HybridHoareLogic}
is focused on the hybrid Hoare logic. Section~\ref{sec:example}
details one step of the parameter identification on the biological
example of the \textit{lacI} repressor interacting with the \textit{Ntr}
system. Finally, we discuss the limits of this approach in
\pref{sec:conclusion}.

%% file: framework.tex
\section{Hybrid Modelling Framework}
\label{sec:framework}

\begin{figure}[t]
  \centerline{
  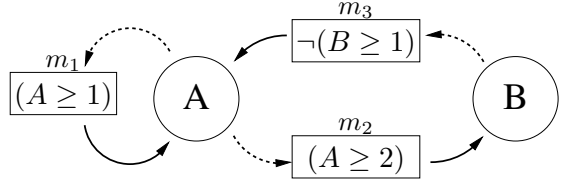
  }  
  \caption{\textbf{The gene network controlling the \textit{lacI}
      repressor regulation} of the lactose operon in \textit{E.~Coli}.
      See \pref{sec:example} for more details.
    \label{fig:InteractionGraph}}
\end{figure}

A gene network is visualised as a labelled directed graph (interaction
graph) in which vertices are either variables (within circles) or
multiplexes (within rectangles), see
Fig.~\ref{fig:InteractionGraph}. Variables abstract genes and their
products, and multiplexes contain propositional formulas that encode
situations in which a variable or a group of variables (inputs of multiplexes,
dashed arrows) influences the evolution of some other variables (output of
multiplexes, plain arrows). A multiplex can encode the formation of
molecular complexes, phosphorylation by a protein, competition of
entities for activation of a promoter, etc.
Definition~\ref{def:network} gives the formal details of a gene network.

\begin{definition}[Hybrid gene regulatory network]\label{def:network}
A hybrid
gene regulatory network (GRN for short) is a tuple
$R=(V,M,E,\mathcal{C})$ where:
\vspace*{-1mm}
\begin{itemize}
\item $V$ is a set whose elements are called \emph{variables} of the
  network. Each variable $v \in V$ is associated with a boundary $b_v\in \N^*$.
\item $M$ is a set whose elements are called \emph{multiplexes}. With each
  multiplex $m\in\!M$ is associated a formula $\varphi_m$
  belonging to the language $\mathcal{L}$ inductively defined by:
  \looseness=-1
\begin{itemize}
  \item If $v\in V$ and $n\in\N$ such that $1\leqslant n
    \leqslant b_v$, then $v\geqslant n$ is an atom of $\mathcal{L}$;
  \item If $\varphi$ and $\psi$ are two formulas of $\mathcal{L}$,
    then $\neg \varphi$, $(\varphi\vee\psi)$, $(\varphi\wedge\psi)$
    and $(\varphi\Rightarrow\psi)$ also belong to $\mathcal{L}$.
\end{itemize}
\item $E$ is a set of edges of the form $(m\rightarrow v)\in M\times V$.
\item $\mathcal{C}=\{C_{v,\omega,n}\}$ is a family of real numbers
  indexed by the tuple $(v,\omega,n)$ where $v$, $\omega$ and $n$
  verify the three following conditions:
  \begin{enumerate}
     \item $v\in V$,
     \item $\omega$ is a subset of $\pred(v)$ where $\pred(v) =
       \{m~|~(m\rightarrow v)\in E\}$, that is, $\omega$ is a set of
       predecessors of $v$,
     \item $n$ is an integer such that $0 \leqslant n \leqslant b_v$.
  \end{enumerate}
  $C_{v,\omega,n}$ is called the \emph{celerity} of $v$ for $\omega$
  at the level $n$.
\end{itemize}
Moreover, celerities are constrained. For each $v\in
V$ and for each $\omega\subset\pred(v)$:
\begin{itemize}
\item either all celerities $C_{v, \omega, n}$ with $0\leqslant
  n\leqslant b_v$ have the same nonzero sign,
\item or there exists $n_0$ such that $C_{v, \omega, n_0} = 0$ and for
  all $n$ such that $0\leqslant n<n_0$, we have $\sgn(C_{v, \omega, n})
  =1$ and for all $n$ such that $n_0<n\leqslant b_v$, we have
  $\sgn(C_{v, \omega, n})=-1$ where the function $\sgn$ is the classical sign function.
\end{itemize}
\end{definition}
Let us remark that the dashed arrows of
Fig.~\ref{fig:InteractionGraph} are not present in the previous
definition. When representing a gene network, it is convenient to
visualise the variables contributing to a particular multiplex, but
from a formal point of view, this information is coded into the
formula of the considered multiplex. 

In the remainder of this section, we focus on the dynamics of such a
gene network. Definition~\ref{def:states} introduces the hybrid
states whereas \pref{def:resources} explains the crucial
notion of resources of a variable at a particular state. 

\begin{definition}[State of a GRN]\label{def:states}
Let $R=(V,M,E,\mathcal{C})$ be a GRN. 
A \emph{hybrid state} of $R$ is a tuple $h=(\eta,\pi)$ where
\begin{itemize}
\item $\eta$ is a function from $V$ to $\N$ such that 
  for all $v\in V$, $0\leqslant\eta(v)\leqslant b_v$;
\item $\pi$ is a function from $V$ to the interval $\Rsegm{0}{1}$ of real
  numbers.
\end{itemize}
$\eta$ is called the \emph{discrete state} of $h$ whereas 
$\pi$ is called its \emph{fractional part}. For simplicity, we note in
the sequel $\eta_v=\eta(v)$ and $\pi_v = \pi(v)$.
We denote $S$ the set of hybrid states of $R$. When there is no
ambiguity, we often use $\eta$ and $\pi$ without explicitly mentioning
$h$. 
\end{definition}

\begin{figure}[t]
  \centerline{
  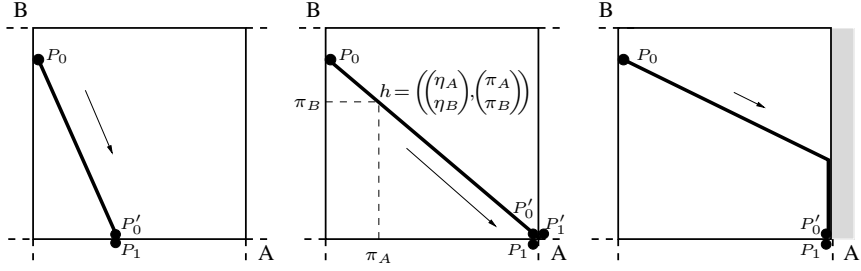
  }  
  \caption{\textbf{Continuous transitions.} 
    Inside each
    state, a continuous transition ($P_0 \rightarrow P_0'$) goes from
    the initial point $P_0$ to the unique point $P'_0$ from which a
    discrete transition takes place ($P_0' \rightarrow P_1$). {\bf
      Left:} The celerity vector allows, without sliding mode, the
    trajectory to directly reach a border which is crossed. {\bf
      Center:} From $P'_0$ two possible discrete transitions can occur:
    $P_0' \rightarrow P_1$ or $P_0' \rightarrow P'_1$. 
    Moreover $(\pi_A , \pi_B)$ corresponds to the fractionnal coordinates of
    a hybrid state $h$ along the path. {\bf Right:} The grey
    zone depicts an external or internal wall. The only possible
    discrete transition is $P_0' \rightarrow P_1$.\label{fig:Transition}}
\end{figure}

Figure~\ref{fig:Transition}-Centre illustrates an example of hybrid
state. The tuple of all fractionnal parts represents coordinates inside the current qualitative state. 

\begin{definition}[Resources]\label{def:resources}
Let $R=(V,M,E,\mathcal{C})$ be a GRN and let $v\in V$. The
satisfaction relation $h\vDash \varphi$, where $h=(\eta,\pi)$ is a
hybrid state of $R$ and $\varphi$ a formula of $\mathcal{L}$, is
inductively defined by:
\begin{itemize}
\item If $\varphi$ is the atom $v\geqslant n$ with $n\in
  \Nsegm{1}{b_v}$, then $h \vDash \varphi$ iff $\eta_v \geqslant n$;
\item If $\varphi$ is of the form $\neg \psi$, then $h\vDash\varphi$
  iff $h \nvDash \psi$;
\item If $\varphi$ is of the form $\psi_1\vee\psi_2$, then $h \vDash
  \varphi$ iff $h \vDash \psi_1$ or $h \vDash \psi_2$ 
  and we proceed similarly for the other connectives.
\end{itemize}
The set of resources of a variable $v$ for a state $h$ is defined by:
$\rho(h,v)=\{m\in \pred(v) \mid h \vDash \varphi_m\}$, that is, the
multiplexes predecessors of $v$ whose formula is satisfied. 
Thus,  $\omega$ is the subset of resources of $v$ iff
the formula $\Phi_v^\omega$ is true:
\begin{eqnarray}\label{eq:formulePhiOmegaV}
\Phi_v^\omega \equiv \Big(\bigwedge_{m \in \omega} \varphi_m\Big) \wedge
\Big(\bigwedge_{m \in R^{-1}(v) \setminus \omega} \neg \varphi_m\Big) \enspace.
\end{eqnarray}
\end{definition}

We note that the evaluation of formula $\Phi_v^\omega$
given in \pref{eq:formulePhiOmegaV}
requires only the valuation of the
discrete state: all hybrid states having the same discrete part have
the same resources. This illustrates the fact that inside a discrete
state, the dynamics is controlled in the same manner, thus the
celerity is the same: the one associated with the current discrete
state and with the current resources $C_{v,\rho(v,\eta),\eta_v}$.
From this celerity, and given a particular hybrid state, one can
compute the touch delay of each variable, which measures the time
necessary for the variable to reach a border of the discrete state.

\begin{definition}[Touch delay]\label{def:touchdelay}
Let $R=(V,M,E,\mathcal{C})$ be a GRN, $v$ be a variable of $V$ and
$h=(\eta,\pi)$ be a hybrid state. We note $\delta_h(v)$ the
\emph{touch delay} of $v$ in $h$ for reaching the border of the
discrete state. More precisely, $\delta_h$ is the function from $V$ to
$\R^+\cup\{+\infty\}$ defined by: 
\begin{itemize}
\item If $C_{v,\rho(v,\eta),\eta_v}=0$ then $\delta_h(v) = +\infty$;
\item If $C_{v,\rho(v,\eta),\eta_v}>0$ then $\delta_h(v) =
  \frac{1-\pi_v}{C_{v,\rho(v,\eta),\eta_v}}$;
\item If $C_{v,\rho(v,\eta),\eta_v}<0$ then $\delta_h(v) =
  \frac{-\pi_v}{C_{v,\rho(v,\eta),\eta_v}}$.
\end{itemize}
\end{definition}

Unfortunately, being at a border of the discrete state
is not sufficient to go beyond the frontier: there may be no other
qualitative level beyond the border (we call such a border an external
wall) or the celerity in the neighbouring state may be of the opposite sign
(internal wall), as given in \pref{def:sliding}.

\begin{definition}[External and internal walls]\label{def:sliding}
Let $R=(V,M,E,C)$ be a GRN, let $v\in V$ be a variable and 
$h=(\eta,\pi)$ a hybrid state. 
\begin{enumerate}
\item $v$ is said to \emph{potentially meet an external wall} if:
  \[\big((C_{v,\rho(v,\eta),\eta_v}<0) \land (\eta_v=0)\big) \lor 
   \big((C_{v,\rho(v,\eta),\eta_v}>0) \land (\eta_v=b_v)\big) \enspace.\]
\item Let $h'=(\eta',\pi')$ be another hybrid state s.t.
  $\eta'_v=\eta_v+\sgn(C_{v,\rho(v,\eta),\eta_v})$ and
  $\eta'_u=\eta_u$ for all $u\neq v$.
  Variable $v$ is said to \emph{potentially meet an internal wall} if:
  \[\sgn(C_{v,\rho(v,\eta),\eta_v})\times
        \sgn(C_{v,\rho(v,\eta'),\eta'_v})=-1 \enspace.\]
\end{enumerate}
We note $sv(h)$ the set of \emph{sliding variables}
that will potentially meet an internal or external wall
in the qualitative state of $h$.
\end{definition}

We note that for all $v \in sv(h)$,
if, in addition, $\delta_h(v)=0$,
then $v$ is located on the said internal wall or external wall.
In this case, its fractional part cannot evolve anymore
in the current qualitative state
(see \pref{fig:Transition}-Right where variable $B$ reaches an external
wall).
We introduce the notion of knocking variables
in \pref{def:knockingVar} which 
are the first variables able to change their discrete levels. 

\begin{definition}[Knocking variables]\label{def:knockingVar}
Let $R=(V, M, E, C)$ be a GRN and $h=(\eta,\pi)$ be a hybrid state.
The set of \emph{knocking variables} is defined by:
  \[first(h)=\{v\in V\setminus sv(h) \mid
    \delta_h(v) \neq +\infty \wedge \forall u\in V\setminus sv(h),
    \delta_h(u) \geqslant \delta_h(v)\} \enspace.\]
Moreover, $\delta_h^{first}$
denotes the time spent in the qualitative state of $h$ when starting
from $h$: for any $x\in first(h)$, $\delta_h^{first} = \delta_h(x)$.
\end{definition}

The set $first(h)$ represents the set of variables whose qualitative
coordinate can change first. If the variable is on an external of
internal wall, it cannot evolve as long as other variables do not
change. Similarly, if the celerity of $v$ in the current state is
null, its qualitative value cannot change because of an infinite touch
delay.

Figure~\ref{fig:Transition} illustrates several evolutions of a gene
network. From a particular hybrid state $P_0$, the dynamics alternates
continuous transitions (within the discrete state) and discrete
transitions (when changing the discrete state). When the trajectory
reaches an external or internal wall (see Right part of the figure)
the variable slides along the wall only if the celerity of some other
variable can drive the trajectory in such a direction. 
This description leads to \pref{def:transition}.

\begin{definition}[Hybrid state space]\label{def:transition}
Let $R=(V,M,E,C)$ be a GRN, we note $\mathcal{R}=(S,T)$ the
\emph{hybrid state space} of $R$ where $S$ is the set of hybrid states
and $T$ is the set of \emph{transitions}: there exists a transition
from state $h'=(\eta',\pi')$ to state $h=(\eta,\pi)$ iff there exists a variable $v\in
first(h')$ such that:
\begin{enumerate}
\item \label{item:defHSP-1} Either $\delta_{h'}(v)\neq 0$ and
  \begin{itemize}
    \item[(i)] $\eta=\eta'$,
    \item[(ii)] $\pi_v=\frac{1+\sgn(C_{v,\rho(v,\eta),\eta_v})}{2}$
      for all $v\in first(h')$,
    \item[(iii)] $\forall u\in V\setminus first(h')$, if $u\notin sv(h')$
      then $\pi_u=\pi'_u+\delta_{h'}(v)\times
      C_{u,\rho(u,\eta'),\eta'_u}$, else 
      $\pi_u=\pi'_u=\frac{1+\sgn(C_{u,\rho(u,\eta'),\eta'_u})}{2}$. 
  \end{itemize}
\item \label{item:defHSP-2} Or $\delta_{h'}(v) = 0$ and
  \begin{itemize}
  \item[(i)] $\eta_v=\eta'_v+\sgn(C_{v,\rho(v,\eta'),\eta'_v})$,
  \item[(ii)] $\pi_v=\frac{1-\sgn(C_{v,\rho(v,\eta'),\eta'_v})}{2}$,
  \item[(iii)] $\forall u\in V\setminus \{v\}$, $\eta_u=\eta'_u$ and
    $\pi'_u=\pi_u$.
  \end{itemize}
\end{enumerate}
\end{definition}

\noindent 
On the one hand, the item \ref{item:defHSP-1}.\ of the definition
describes the \emph{continuous transitions}: the transitions lead to
the last hybrid state inside the current discrete state which could
give rise to a qualitative change.  On the other hand, the item
\ref{item:defHSP-2}.\ describes the \emph{discrete transitions}: if
the system cannot evolve anymore within the current discrete state,
the trajectory goes through a border.

Let us remark that, if $v$ does not encounter any internal or
external wall, the time spent in a particular discrete state 
is computed by the formula
$\Delta t = \frac{\pi_v - \pi'_v}{C_{v, \omega, n}}$ 
where $\pi'_v$ (\resp $\pi_v$) is the fractional part of the variable
$v$ at the entrance (\resp exit) of the current state and
$C_{v, \omega, n}$ is the celerity of the variable $v$ inside the
current discrete state. $\Delta t$ is called the \emph{duration} of the
continuous transition.

%% file: interaction_graph_example.pdf_t
\begin{picture}(0,0)%
\includegraphics{./interaction_graph_example.pdf}%
\end{picture}%
\setlength{\unitlength}{2901sp}%
\begingroup\makeatletter\ifx\SetFigFont\undefined%
\gdef\SetFigFont#1#2#3#4#5{%
  \reset@font\fontsize{#1}{#2pt}%
  \fontfamily{#3}\fontseries{#4}\fontshape{#5}%
  \selectfont}%
\fi\endgroup%
\begin{picture}(4655,1590)(-461,-973)
\put(  1,-331){\makebox(0,0)[b]{\smash{{\SetFigFont{11}{13.2}{\rmdefault}{\mddefault}{\updefault}{\color[rgb]{0,0,0}$(A \geq 1)$}%
}}}}
\put(  1,-16){\makebox(0,0)[b]{\smash{{\SetFigFont{10}{12.0}{\rmdefault}{\mddefault}{\updefault}{\color[rgb]{0,0,0}$m_1$}%
}}}}
\put(2476,119){\makebox(0,0)[b]{\smash{{\SetFigFont{11}{13.2}{\rmdefault}{\mddefault}{\updefault}{\color[rgb]{0,0,0}$\neg (B \geq 1)$}%
}}}}
\put(2476,434){\makebox(0,0)[b]{\smash{{\SetFigFont{10}{12.0}{\rmdefault}{\mddefault}{\updefault}{\color[rgb]{0,0,0}$m_3$}%
}}}}
\put(2476,-556){\makebox(0,0)[b]{\smash{{\SetFigFont{10}{12.0}{\rmdefault}{\mddefault}{\updefault}{\color[rgb]{0,0,0}$m_2$}%
}}}}
\put(2476,-871){\makebox(0,0)[b]{\smash{{\SetFigFont{11}{13.2}{\rmdefault}{\mddefault}{\updefault}{\color[rgb]{0,0,0}$(A \geq 2)$}%
}}}}
\end{picture}%

%% file: transition_examples.pdf_t
\begin{picture}(0,0)%
\includegraphics{./transition_examples.pdf}%
\end{picture}%
\setlength{\unitlength}{2901sp}%
\begingroup\makeatletter\ifx\SetFigFont\undefined%
\gdef\SetFigFont#1#2#3#4#5{%
  \reset@font\fontsize{#1}{#2pt}%
  \fontfamily{#3}\fontseries{#4}\fontshape{#5}%
  \selectfont}%
\fi\endgroup%
\begin{picture}(7255,2287)(1329,-2333)
\put(2311,-1996){\makebox(0,0)[lb]{\smash{{\SetFigFont{6}{7.2}{\rmdefault}{\mddefault}{\updefault}{\color[rgb]{0,0,0}$P_0'$}%
}}}}
\put(1681,-556){\makebox(0,0)[lb]{\smash{{\SetFigFont{6}{7.2}{\rmdefault}{\mddefault}{\updefault}{\color[rgb]{0,0,0}$P_0$}%
}}}}
\put(2311,-2221){\makebox(0,0)[lb]{\smash{{\SetFigFont{6}{7.2}{\rmdefault}{\mddefault}{\updefault}{\color[rgb]{0,0,0}$P_1$}%
}}}}
\put(4156,-556){\makebox(0,0)[lb]{\smash{{\SetFigFont{6}{7.2}{\rmdefault}{\mddefault}{\updefault}{\color[rgb]{0,0,0}$P_0$}%
}}}}
\put(6676,-556){\makebox(0,0)[lb]{\smash{{\SetFigFont{6}{7.2}{\rmdefault}{\mddefault}{\updefault}{\color[rgb]{0,0,0}$P_0$}%
}}}}
\put(5896,-1996){\makebox(0,0)[lb]{\smash{{\SetFigFont{6}{7.2}{\rmdefault}{\mddefault}{\updefault}{\color[rgb]{0,0,0}$P_1'$}%
}}}}
\put(8056,-1996){\makebox(0,0)[lb]{\smash{{\SetFigFont{6}{7.2}{\rmdefault}{\mddefault}{\updefault}{\color[rgb]{0,0,0}$P_0'$}%
}}}}
\put(8056,-2221){\makebox(0,0)[lb]{\smash{{\SetFigFont{6}{7.2}{\rmdefault}{\mddefault}{\updefault}{\color[rgb]{0,0,0}$P_1$}%
}}}}
\put(5581,-2221){\makebox(0,0)[lb]{\smash{{\SetFigFont{6}{7.2}{\rmdefault}{\mddefault}{\updefault}{\color[rgb]{0,0,0}$P_1$}%
}}}}
\put(5626,-1861){\makebox(0,0)[lb]{\smash{{\SetFigFont{6}{7.2}{\rmdefault}{\mddefault}{\updefault}{\color[rgb]{0,0,0}$P_0'$}%
}}}}
\put(4501,-871){\makebox(0,0)[lb]{\smash{{\SetFigFont{7}{8.4}{\rmdefault}{\mddefault}{\updefault}{\color[rgb]{0,0,0}$h\!=\!\left(\!\!
\left(\!\begin{array}{c} \!\eta_A \\ \!\eta_B \end{array}\!\!\right)\!\!,\!\! 
\left(\!\begin{array}{c} \!\pi_A  \\ \!\pi_B  \end{array}\!\!\right)
\!\!\right)
$}%
}}}}
\put(3781,-961){\makebox(0,0)[lb]{\smash{{\SetFigFont{7}{8.4}{\rmdefault}{\mddefault}{\updefault}{\color[rgb]{0,0,0}$\pi_B$}%
}}}}
\put(4501,-2266){\makebox(0,0)[b]{\smash{{\SetFigFont{7}{8.4}{\rmdefault}{\mddefault}{\updefault}{\color[rgb]{0,0,0}$\pi_A$}%
}}}}
\end{picture}%

%% file: hoare12.tex
\section{Hybrid Hoare Logic}
\label{sec:HybridHoareLogic}
\subsection{Languages}

This section is dedicated to the presentation of the Hoare logic
adapted to our hybrid formalism.
We first detail the notion of terms (\pref{def:terms})
in order to define the property language (\pref{def:property})
later used for pre- and postconditions,
the assertion language (\pref{def:assert})
and its semantics (\pref{def:assertsem})
used to describe biological knowledge on a hybrid transition,
and the path language (\pref{def:path})
later used to describe an observed trace inside a hybrid gene regulatory network.

\begin{definition}[Terms]
\label{def:terms}
The \emph{terms} are inductively defined as follows:
\begin{itemize}
  \item The variables $\eta_u$, $\pi_u$ and $\pi'_u$  where $u\in V$ are terms;
  \item The variables $C_{u, \omega, n}$ where $u\in V$, $\omega
    \subset \pred(u)$ and $n\in \Nsegm{0}{b_u}$ are terms;
  \item The variables and constants of $\mathbb{R}$ and $\mathbb{N}$ are terms;
  \item The set of considered connectives that create new terms are:
    $+$, $-$, $\times$, $/$.
\end{itemize}
\end{definition}

We denote by $\square$ any of the following symbols: $<$, $\leq$,
$>$, $\geq$, $=$, $\not =$.
Moreover we note $Terms_d$ (\resp $Terms_h$) the set of
“discrete” terms (\resp “hybrid” terms) built on variables 
$\eta_u$ and on variables and constants of $\mathbb{N}$
(\resp on variables $\pi_u$,
$\pi'_u, C_{u, \omega, n}$ and on
variables and constants of $\mathbb{R}$).

\begin{definition}[Property language ${\cal L}_C$]
\label{def:property}
The atoms of the \emph{property language} are of two types:
    \begin{itemize}
      \item \emph{Discrete atoms} are of the form: 
        $n\ \Box\ n'$  where $n, n' \in Terms_d$;
      \item \emph{Hybrid atoms} are of the form: 
        $f\ \Box\ f'$  where $f, f' \in Terms_h$;
    \end{itemize}

The \emph{discrete conditions} are defined by:
$D ~~ :== ~~ a_d ~|~ \lnot D  ~|~  D \land D ~|~  D \lor D $
where $a_d$ is a discrete atom.

The \emph{hybrid conditions} are defined by:
$H ~~ :== ~~ a_d ~|~ a_h ~|~ \lnot H  ~|~  H \land H ~|~  H \lor H $
where $a_d$ and $a_h$ are respectively a discrete atom and a hybrid
one.

A \emph{property} is a couple $(D, H)$
formed by a discrete and a hybrid condition.
All properties form the \emph{property language} ${\cal L}_C$.
\end{definition}

A hybrid state $h$ satisfies a property $\varphi = (D, H) \in {\cal L}_C$
iff both $D$ and $H$ hold in $h$,
by using the usual meaning of the connectives;
in this case, we note $h \models \varphi$.

\begin{definition}[Assertion language ${\cal L}_A$]
\label{def:assert}
The \emph{assertion language} ${\cal L}_A$ is defined by the following grammar:
\[a ~~ :== ~~ \top ~|~ C_v ~\square~ c ~|~ slide(v) ~|~ slide^+(v)
~|~ slide^-(v) ~|~ \neg a ~|~ a \land a ~|~ a \lor a\]
where $v\in V$ is a variable name and $c \in \R$ is a real number. 
\end{definition}

\begin{definition}[Semantics of the assertion couple $(\Delta t,a)$] 
\label{def:assertsem}
Let us consider a hybrid state $h'=(\eta,\pi')$ and the unique continuous
transition starting from $h'$ and ending in $h=(\eta,\pi)$. 
The satisfaction relation between the continuous transition 
$h'\longrightarrow h$ and an assertion couple $(\Delta t,a) \in \R^+ \times
{\cal L}_A$ is noted $(h',h) \models (\Delta t,a)$, by overloading of notation, and is
defined as follows:
\begin{itemize}
\item If $a\equiv \top$, $(h',h)\models (\Delta t,a)$ 
  iff $\delta_{h'}^{first} = \Delta t$.
\item If $a$ is of the form $(C_u ~\square~ c)$, $(h',h)\models (\Delta t,a)$
  iff $\delta_{h'}^{first} = \Delta t$ and $(C_{u,\rho(u,\eta),\eta_u}
  ~\square~ c)$.
\item If $a$ is of the form $slide(v)$,
  $(h',h)\models (\Delta t,a)$ iff
  $\delta_{h'}^{first} = \Delta t$ and 
  $\delta_{h'}(v) <~\delta_{h'}^{first}$.
\item If $a$ is of the form $slide^+ (v)$ (\resp $slide^-(v)$),
  $(h',h)\models (\Delta t,a)$ iff
  $\delta_{h'}^{first} = \Delta t$ and
  $\delta_{h'}(v) < \delta_{h'}^{first}$ and
  $C_{v,\rho(v,\eta),\eta_v} > 0$
  (\resp $C_{v,\rho(v,\eta),\eta_v} < 0$).
\item If $a$ is of the form $\neg a'$,  
  $(h',h)\models (\Delta t,a)$
    iff $\delta_{h'}^{first} = \Delta t$ and $(h',h)\not\models (\Delta t,a')$.
\item If $a$ is of the form $a' \land a''$ (\resp $a' \lor a''$),  
  $(h',h)\models (\Delta t,a)$
    iff $(h',h)\models (\Delta t,a')$ and (\resp or) $(h',h)\models (\Delta t,a'')$.
\end{itemize}
\end{definition}

\begin{definition}[Path language ${\cal L}_P$]
\label{def:path}
The \emph{(discrete) path atoms} are defined~by:
$dpa ~~ :== ~~ v+ ~|~ v-$
where $v\in V$ is a variable name.

The \emph{(discrete) paths} are  defined by:
$~~p :== ~ \emptyseq ~|~ (\Delta t,assert,dpa) ~|~ p ~;~ p$
where $\Delta t$ is either a real constant or a variable, $assert$ is an
assertion, and $dpa$ is a discrete path atom.   
\end{definition}

\subsection{Hoare Triples}

Using the languages defined in the previous section,
we give here the syntax (\pref{def:triple})
and the semantics (\pref{def:tripleSemantics})
of a Hoare triple
in the scope of our hybrid formalism,
which are natural extensions of the classical definitions.

\begin{definition}[Hybrid Hoare Triples]\label{def:triple}
A Hoare triple for a given GRN is an expression of the form
$\hoare{Pre}{p}{Post}$
where $Pre$ and $Post$, called
\emph{precondition} and \emph{postcondition} respectively,
are properties of ${\cal L}_C$,
and $p$ is a path from ${\cal L}_P$.
\end{definition}

\begin{definition}[Semantics of Hoare triples]\label{def:tripleSemantics}
  We say that a Hoare triple $\hoare{Pre}{p}{Post}$ is \emph{satisfied}
  if:
  \begin{itemize}
    \item If $p$ is atomic and is of the form 
      $(\Delta t, assert, v+)$
      (\resp $(\Delta t, assert, v-)$)
      then for all $h'_1=(\eta_1, \pi'_1) \models Pre$,
      there exists two hybrid states $h_1=(\eta_1,\pi_1)$ and
      $h'_2=(\eta_2,\pi'_2)$
      with $\etai{v}{2} = \etai{v}{1} + 1$
      (\resp $\etai{v}{2} = \etai{v}{1} - 1$)
      such that:
      \begin{itemize}
        \item there exists a continuous transition 
          from $h'_1$ towards $h_1$ 
          such that $(h'_1,h_1) \models (\Delta t,assert)$,
        \item  there exists a discrete transition from $h_1$
          towards $h'_2$,
        \item $h'_2 \models Post$;
      \end{itemize}
    \item If $p = p_1 ; p_2$ is a sequence of paths,
      then there exist  $ Post_1$ and $Pre_2$ of the property language
      such that both triples $\hoare{Pre}{p_1}{Post_1}$ and $\hoare{Pre_2}{p_2}{Post}$
      are satisfied and $Post_1 \Rightarrow Pre_2$.
  \end{itemize}
\end{definition}

\begin{figure}[t]
  \centerline{
  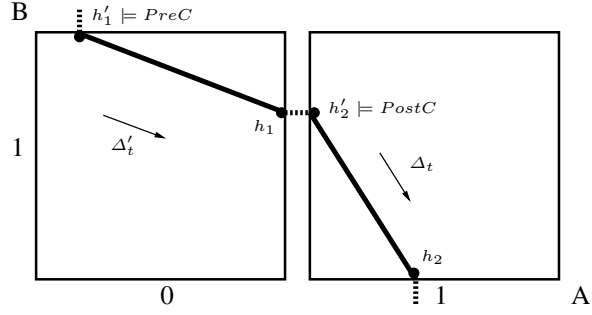
  }  
  \caption{\label{fig:pathExample}
  \textbf{Hoare triple example: $\hoare{PreC}{(\Delta_t', \top, A+)}{PostC}$.} 
  Starting from the hybrid state $h_1'$, and considering the path in bold line,
  it is possible to chain a hybrid transition
  of duration $\Delta_t'$ ($h_1' \rightarrow h_1$)
  and a discrete transition ($h_1 \rightarrow h_2'$)
  so that the discrete level of $A$ is increased:
  this Hoare triple is satisfied.}
\end{figure}

Figure~\ref{fig:pathExample} represents an execution inside a GRN containing
several discrete transitions.
It illustrates the fact that both
the precondition and postcondition of a particular Hoare triple are associated
with the hybrid states corresponding to the entrance in the related discrete states.

%% file: one_path_example.pdf_t
\begin{picture}(0,0)%
\includegraphics{./one_path_example.pdf}%
\end{picture}%
\setlength{\unitlength}{4144sp}%
\begingroup\makeatletter\ifx\SetFigFont\undefined%
\gdef\SetFigFont#1#2#3#4#5{%
  \reset@font\fontsize{#1}{#2pt}%
  \fontfamily{#3}\fontseries{#4}\fontshape{#5}%
  \selectfont}%
\fi\endgroup%
\begin{picture}(3498,1853)(1381,-1314)
\put(1985,-360){\makebox(0,0)[lb]{\smash{{\SetFigFont{6}{7.2}{\rmdefault}{\mddefault}{\updefault}{\color[rgb]{0,0,0}$\Delta_t'$}%
}}}}
\put(3756,-470){\makebox(0,0)[lb]{\smash{{\SetFigFont{6}{7.2}{\rmdefault}{\mddefault}{\updefault}{\color[rgb]{0,0,0}$\Delta_t$}%
}}}}
\put(2833,-249){\makebox(0,0)[lb]{\smash{{\SetFigFont{6}{7.2}{\rmdefault}{\mddefault}{\updefault}{\color[rgb]{0,0,0}$h_1$}%
}}}}
\put(3829,-1023){\makebox(0,0)[lb]{\smash{{\SetFigFont{6}{7.2}{\rmdefault}{\mddefault}{\updefault}{\color[rgb]{0,0,0}$h_2$}%
}}}}
\put(3276,-140){\makebox(0,0)[lb]{\smash{{\SetFigFont{6}{7.2}{\rmdefault}{\mddefault}{\updefault}{\color[rgb]{0,0,0}$h_2' \models PostC$}%
}}}}
\put(1875,413){\makebox(0,0)[lb]{\smash{{\SetFigFont{6}{7.2}{\rmdefault}{\mddefault}{\updefault}{\color[rgb]{0,0,0}$h_1' \models PreC$}%
}}}}
\end{picture}%

%% file: hoare3.tex
\subsection{Weakest Precondition Calculus}
\label{ssec:wp}

In this section, we detail the computation of the weakest precondition
related to a given program and postcondition.
This method is inspired from the original weakest precondition
proposed by Dijkstra~\cite{Dijkstra:1975:GCN:360933.360975}
with several additions allowing to take the hybrid dynamics into account.
Indeed, given the semantics of the hybrid formalism defined in \pref{sec:framework},
several conditions have to be met to allow the execution
of a $v+$ or $v-$ instruction:
amongst others,
the variable $v$ must be the first to be able to cross a border,
the celerities must allow the related continuous and discrete transitions,
there is a relationship between the fractional parts of states before and after 
a discrete transition...

In \pref{def:wp}, we give a formal definition of the weakest precondition
$(D', H')$
of a path program according to a given postcondition $(D, H)$.
The interesting terminal cases are the increment ($v+$)
and decrement ($v-$) instructions.

\begin{definition}[Weakest precondition]
  \label{def:wp}
  Let $p$ be a path program and $Post = \propinline{D}{H_f}$ a property
  which will have the role of a post-condition
  parameterized by a final state index $f$.
  The \emph{weakest precondition} attributed to $p$ and $Post$
  is a property:
  $\WP^i_f(p, Post) \equiv \propinline{D'}{H'_{i,f}}$,
  parameterized by a fresh initial state index $i$
  and the same final state $f$,
  and whose value is recursively defined by:
  \begin{itemize}
    \item If $p = \emptyseq$ is the empty sequence program,
      then $D' \equiv D$ and $H'_{i,f} \equiv H_f$;
    \item If $p = (\Delta t, assert, v+)$ is an atom, with $v \in V$:
      \begin{itemize}
        \item $D' \equiv D[\eta_v \backslash \eta_v + 1]$,
        \item $H'_{i,f} \equiv H_f \wedge \Phi_v^+(\Delta t) \wedge \F_v(\Delta t) \wedge
          \neg \W^+_v \wedge \A(\Delta t, assert) \wedge \Rep_v$;
      \end{itemize}
    \item If $p = (\Delta t, assert, v-)$ is an atom, with $v \in V$:
      \begin{itemize}
        \item $D' \equiv D[\eta_v \backslash \eta_v - 1]$,
        \item $H'_{i,f} \equiv H_f \wedge \Phi_v^-(\Delta t) \wedge \F_v(\Delta t) \wedge
          \neg \W^-_v \wedge \A(\Delta t, assert) \wedge \Rep_v$;
      \end{itemize}
    \item If $p = p_1 ; p_2$ is a concatenation of programs:
      \[\WP^i_f(p_1 ; p_2, Post) \equiv \WP^i_m(p_1, \WP^m_f(p_2, Post))\]
      which is parameterized by a fresh intermediate state index $m$;
  \end{itemize}
  where $\Phi_v^+(\Delta t)$, $\Phi_v^-(\Delta t)$, $\W_v^+$, $\W_v^-$, $\F_v(\Delta t)$, $\A(\Delta t, assert)$ and $\Rep_v$ 
  are sub-properties given in \pref{ap:subformulas}.
\end{definition}

In the following, we informally describe the content of the sub-properties
that are used in the weakest precondition construction.
The complete formal definitions are given in \pref{ap:subformulas}.
It has to be noted that all of these properties implicitly depend on the
indices $i$ and $f$ used in \pref{def:wp}.
In the following, $v \in V$ is a component and $\Delta t \in \sR^+$ is a time delay.

\begin{itemize}
  \item $\Phi_v^+(\Delta t)$ (\resp $\Phi_v^-(\Delta t)$)
    describes the conditions in which
    $v$ increases (\resp decreases) its discrete expression level:
    its celerity in the current state must be positive (\resp negative),
    and its final fractional part $\pi_v$ 
    depends on $\Delta t$.
  \item $\W^+_v$ (\resp $\W^-_v$)
    states that there is an internal or external wall
    preventing $v$ from increasing (\resp decreasing) its qualitative state;
    this sub-property must of course be false for the discrete transition to take place.
  \item $\F_v(\Delta t)$
    states that $v$ belongs to the set of knocking variables,
    that is, the variables which can first
    change their qualitative state;
    in other words, all components other than $v$ must either
    reach their border after $v$, or face an internal or external wall.
  \item $\A(\Delta t, assert)$
    translates all assertion symbols given in $assert$
    expressing constraints on celerities and slides
    into property language.
  \item $\Rep_v$
    establishes a junction between the fractional parts of two successive states
    linked by a discrete transition:
    it states that all fractional parts are left the same, except for
    the variable $v$ performing the transition,
    whose fractional part swaps between $0$ and $1$.
\end{itemize}

\subsection{Backward strategy}
\label{ssec:backward}

A Hoare triple $\hoare{Pre}{path}{Post}$ being given,
we call backward strategy the proof strategy defined inductively on
$path$ as follows:
\begin{enumerate}
  \item If $path$ is of the form $(\Delta t,assert,v+)$, $(\Delta t,assert,v-)$
    or $\emptyseq$, then compute the precondition of $path$;
  \item If $path$ is of the form $p_1 ; p_2$
    where $p_2$ is of the form $(\Delta t,assert,v+)$ or $(\Delta t,assert,v-)$,
    then compute the precondition just before $p_2$ and iterate for path $p_1$.
\end{enumerate}
Notice that, these two items being mutually exclusive, the backward strategy
generates a unique proof tree.
Furthermore,
given the semantics of Hoare triples, every path can be seen as completely 
linear, justifying the backward strategy.

%% file: example.tex
\section{Example: Controlling the \textit{lacI} Repressor by \textit{NRIp}}
\label{sec:example}

\subsection{Presentation}

We focus on the \textit{lacI} repressor regulation of the lactose 
operon in \textit{E.~Coli}. An operon is a functional DNA unity which 
regroups some genes in order to transcribe them one after the other. These 
genes are regulated by only one promoter allowing a coordinated control 
of their synthesis. In the case of the lactose operon, three genes are controlled: 
lacZ, lacY and lacA which produce the proteins $\beta$-galactosidase, 
permease and thiogalactoside transacetylase, respectively. The 
$\beta$-galactosidase cleaves lactose into glucose and galactose. The 
permease permits the passive transport of the lactose. Finally, the 
thiogalactoside transacetylase confers the detoxification of the 
cell~\cite{andrews1976thiogalactoside}.

The lactose operon is controlled by two sites found ahead of the genes: 
the \textit{lac~p} promoter and the \textit{lac~o} operator. The RNA polymerase 
binds to the promoter and activates the operon. On the contrary, the \textit{lacI} repressor 
binds to the operator and prevents the transcription of the operon genes. In 
presence of lactose, \textit{lacI} repressing activity is suppressed, allowing 
the breaking up of the lactose.

Atkinson \textit{et al.}~\cite{atkinson2003development} used the
lactose operon and the \textit{Ntr} systems to enable oscillatory
behaviours (\pref{fig:InteractionGraph}). Vertex $A$ encodes the
\textit{NRI} protein due to the \textit{glnG} gene combined with the
\textit{glnA} promoter. This promoter is regulated by the
phosphorylated \textit{NRI} (\textit{NRIp}; self-loop on $A$) and by
\textit{lacI} (activation from vertex $B$). Finally, the \textit{lacI}
gene repressor, fused with \textit{glnK} promoter (vertex $B$), is regulated
by \textit{NRIp}.

Because the \textit{glnK} promoter is activated at high \textit{NRIp} 
levels~\cite{atkinson2003development}, we assumed that the auto-activation 
of $A$ is necessary before regulating $B$. In other words, the quantity of 
\textit{NRIp} should be enough to stimulate \textit{lacI}, which explains 
the quantitative thresholds indicated into the interaction graph for 
vertex $A$. Similarly, we only have one threshold crossed for vertex $B$:
$\neg (B \geq 1)$, because \textit{lacI} only negatively interacts with the 
\textit{NRI} gene. A comparative study of the behaviour of $\beta$-galactosidase in 
modelling predictions and in experimental conditions showed same sustained 
oscillation~\cite{purcell2010comparative}.

These oscillations are known to pass through the following sequence of discrete states:
$(A=2,B=0)$, $(2,1)$, $(1,1)$, $(1,0)$, $(2,0)$. Thus, we consider the following 
Hoare triple (for better readability, tuples are written vertically):


\[
  \prop{D_4}{H_4}
  \patha{T_4}{\top}{B+}
    \stackedconcat{D_3, H_3}
  \patha{T_3}{\slide^+(B)}{A-}
    \stackedconcat{D_2, H_2}
  \patha{T_2}{\top}{B-}
    \stackedconcat{D_1, H_1}
  \patha{T_1}{\top}{A+}
  \prop{D_0 \equiv (\eta_A=2 \wedge \eta_B=0)}{H_0 \equiv \top}
    \enspace.
\]

Intuitively, when starting from a state satisfying the discrete part $D_4$ and the hybrid part $H_4$,
 the system evolves until reaching the state $\eta_A = 2 \wedge \eta_B=0$ 
($D_0$) and satisfying the hybrid part $H_0$. From the initial state, \textit{lacI} is 
synthesised due to the activation of its gene (transition $B+$), then this 
repressor inhibits the gene producing the \textit{NRI} protein (transition $A-$).
Then, the gene of \textit{lacI} doesn't remain active (transition $B-$)
which permits the activation of the gene of the \textit{NRI} 
protein (transition $A+$), consequently reaching the final state. 

\begin{figure}[t]
  \centerline{
  \scalebox{.65}{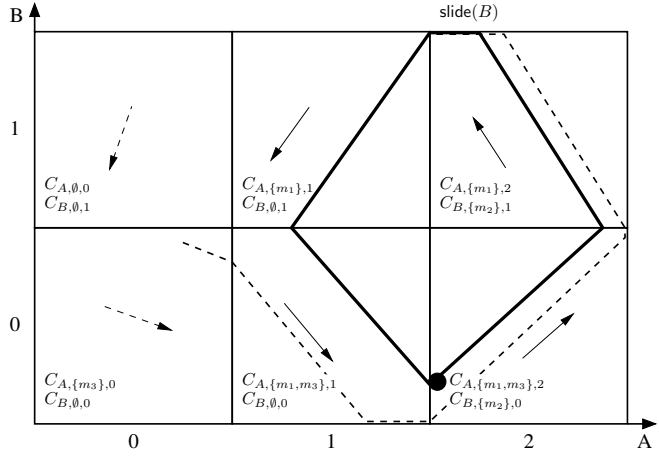}
  }
  \caption{\textbf{State graph of the lacI/Ntr system.} Here, we suppose 
  that only the sign of celerities is known. 
  The shape of the limit cycle depends on the celerities into each crossed 
  state and on biologically known slides. The black circle is the initial state 
  we used. The dashed lines show that there exist other initial states
  that do not belong to the limit cycle but whose trajectories end into it.
  \label{fig:resultExample}}
\end{figure}

We also assumed that the activation of the repressor reached its maximum 
concentration ($\slide(B)$, see \pref{fig:resultExample}) to act on the 
lactose operon when \textit{lacI} is present. Finally, the lack of temporal 
data prevented us to assign a value to the duration $\Delta t$ of each phase,
therefore we used the constants $T_1$, $T_2$, $T_3$ and $T_4$ for these terms.

Using the backward strategy of \pref{ssec:backward} on this Hoare triple allows the determination of each intermediate 
property $(D_i,H_i)$. Intuitively, we would like to represent a cyclic behaviour 
as in \pref{fig:resultExample}, which would lead to make identical the pre- 
and postconditions. We first compute the weakest precondition $(D_4,H_4)$ of 
the previous Hoare triple.

\subsection{First Step of the Backward Strategy}
\label{ssec:step1example}

In this section,
we focus on the first step of the backward strategy, corresponding to
the following Hoare triple:
\[
  \prop{D_1}{H_1}
  \patha{T_1}{\top}{A+}
  \prop{D_0 \equiv (\eta_A=2 \wedge \eta_B=0)}{H_0 \equiv \top}
    \enspace.
\]
Furthermore,
we only show the final result of each sub-formula
of the weakest precondition in order to give
the reader a hint of the results obtained.
Indeed, a lot of simplifications can be made inside these sub-formulas
because they depend on the discrete state of each step,
which is always fully known in this example,
or by using some particularities of our hybrid formalism.
The details of the computation for this first step
are given in \pref{ap:step1}.

First of all, \pref{def:wp} gives:
\begin{align}
  D_1 & \equiv D_0 [\eta_A \backslash \eta_A+1] \equiv (\eta_A+1=2) \wedge (\eta_B=0) \equiv 
    (\eta_A=1) \wedge (\eta_B=0) \enspace, \\
  H_1 & \equiv H_0 \wedge \Phi_A^+(T_1) \wedge \F_A(T_1)
    \wedge \neg \W^+_A \wedge \A(T_1, assert) \wedge \Rep_A \enspace.
\end{align}
The final value of the interesting sub-formulas are given in the following:
\begin{equation}\label{eq:phi}
\Phi_A^+(T_1) \equiv (\pii{A}{1} = 1) \wedge (C_{A,\{m_1,m_3\},1} > 0) 
  \wedge (\pii{A}{1}' = \pii{A}{1} - C_{A,\{m_1,m_3\},1} \cdot T_1) \enspace,
\end{equation}
\begin{equation}\label{eq:first}
\F_A(T_1) \equiv \neg (C_{B,\emptyset,0} > 0) \vee
  \neg (\pii{B}{1}' > \pii{B}{1} - C_{B,\emptyset, 0} \cdot T_1) \enspace,
\end{equation}
\begin{equation}\label{eq:junc}
\Rep_A \equiv (\pii{B}{1} = \pii{B}{0}') \wedge (\pii{A}{1} = 1-\pii{A}{0}') \enspace.
\end{equation}

Here, \pref{eq:phi} gives some requirements for $A$ to increase its discrete level:
its final fractional part ($\pii{A}{1}$) is on border $1$,
its celerity in the current discrete state is positive,
and its initial fractional part is directly given by the time $T_1$ spent in this state.
Moreover, \pref{eq:first} states that,
because $B$ cannot cross a discrete threshold before $A$,
then it must face a wall, or be too far away from its border to cross it.
It is required that $A$ does not meet an internal (or external) wall in this example,
which is always the case here because $\neg \W^+_A \equiv \top$.
Furthermore, for this particular path,
we have no assertion constraining the continuous transition, thus:
$\A(T_1, assert) \equiv \top$.
Finally, \pref{eq:junc} links the fractional part
of both variables in the current discrete state with
their values in the next one.
These results lead to the final value of the hybrid part $H_1$:
\begin{equation}\label{eq:ExampleResult}
\begin{aligned}
  H_1 &\equiv \Big( \neg 
  (C_{B,\emptyset,0} > 0) \vee \neg (\pii{B}{1}' > \pii{B}{0}' - 
  C_{B,\emptyset,0} \cdot T_1) \Big)\\
  &\wedge (C_{A,\{m_1,m_3\},1} > 0) \wedge (\pii{A}{1}' = 1 - 
  C_{A,\{m_1,m_3\},1} \cdot T_1) \wedge (\pii{A}{0}' = 0) \enspace.
\end{aligned}
\end{equation}
Therefore, we obtained constraints for each celerity as well as each fractional 
part in this current state.

\subsection{Final Result}

Using the backward strategy,
we calculated the constraints of the discrete part $D_4$ and the hybrid part 
$H_4$ step by step (all details are given in
\pref{ap:step2}, \ref{ap:step3} and \ref{ap:step4}).
We also took into account the knowledge of the limit cycle
in order to add another particular constraint
(see \pref{ap:limitCycle}).
The final result is the following:
\begin{equation}
\begin{aligned}
  H_F &\equiv \big( \neg (C_{B,\emptyset,0} > 0) \vee \neg (1 > \pii{B}{0}'
   - C_{B,\emptyset,0} \cdot T_1) \big)\\
  &\quad \wedge (C_{A,\{m_1,m_3\},1} > 0) \wedge (\pii{A}{1}' = 1 - 
  C_{A,\{m_1,m_3\},1} \cdot T_1) \\
  &\quad \wedge \big( \neg (C_{A,\{m_1\},1}>0) \vee \neg (1 > \pii{A}{1}' - 
  C_{A,\{m_1\},1} \cdot T_2) \big) \\
  &\quad \wedge \big((C_{A,\emptyset,0} > 0) \vee \neg (C_{A,\{m_1\},1}<0) \vee 
  \neg (1 < \pii{A}{1}' - C_{A,\{m_1\},1} \cdot T_2) \big) \\
  &\quad \wedge (C_{B,\emptyset,1} < 0) \wedge (1 = 0 - C_{B,\emptyset,1} \cdot 
  T_2) \\
  &\quad \wedge \big( \neg (C_{B,\{m_2\},1}<0) \vee \neg (0 < 1 - C_{B,\{m_2\},1} 
  \cdot T_3) \big)\\
  &\quad \wedge \big( C_{A,\{m_1\},2} < 0 \wedge (\pii{A}{3}' = 0 - C_{A,\{m_1\},2} 
  \cdot T_3) \big) \\
  &\quad \wedge \big( \neg (C_{B,\{m_2\},1} > 0) \vee (0 > 1 - C_{B,\{m_2\},1} 
  \cdot T_3) \big)\\
  &\quad \wedge \big( \neg (C_{A,\{m_1,m_3\},2}<0) \vee \neg (0 < \pii{A}{3}' - 
  C_{A,\{m_1,m_3\},2} \cdot T_4) \big) \\
  &\quad \wedge (C_{B,\{m_2\},0} > 0) \wedge (\pii{B}{0}' = 1 - C_{B,\{m_2\},0} 
  \cdot T_4) \enspace.
\end{aligned}
\label{eq:FinalResult}
\end{equation}

Each celerity included in this biological system is mentioned in a constraint 
depending on the temporal information and on the fractional parts of the 
corresponding variable inside the related discrete state. The constraints are 
established for the limit cycle. A constraint solver will then be useful to 
reduce the formulas and identify the sets of exact values
which can be attributed to the celerities.
Once a suitable set of parameters is chosen,
a comparison between the simulation of this model
and biological experiments should be carried out
to assess if the behaviours are similar.

%% file: result_example.pdf_t
\begin{picture}(0,0)%
\includegraphics{./result_example.pdf}%
\end{picture}%
\setlength{\unitlength}{4144sp}%
\begingroup\makeatletter\ifx\SetFigFont\undefined%
\gdef\SetFigFont#1#2#3#4#5{%
  \reset@font\fontsize{#1}{#2pt}%
  \fontfamily{#3}\fontseries{#4}\fontshape{#5}%
  \selectfont}%
\fi\endgroup%
\begin{picture}(5932,4132)(571,-3541)
\put(901,-3121){\makebox(0,0)[lb]{\smash{{\SetFigFont{10}{12.0}{\rmdefault}{\mddefault}{\updefault}{\color[rgb]{0,0,0}$C_{B,\emptyset,0}$}%
}}}}
\put(901,-2941){\makebox(0,0)[lb]{\smash{{\SetFigFont{10}{12.0}{\rmdefault}{\mddefault}{\updefault}{\color[rgb]{0,0,0}$C_{A,\{m_3\},0}$}%
}}}}
\put(901,-1321){\makebox(0,0)[lb]{\smash{{\SetFigFont{10}{12.0}{\rmdefault}{\mddefault}{\updefault}{\color[rgb]{0,0,0}$C_{B,\emptyset,1}$}%
}}}}
\put(901,-1141){\makebox(0,0)[lb]{\smash{{\SetFigFont{10}{12.0}{\rmdefault}{\mddefault}{\updefault}{\color[rgb]{0,0,0}$C_{A,\emptyset,0}$}%
}}}}
\put(2701,-1141){\makebox(0,0)[lb]{\smash{{\SetFigFont{10}{12.0}{\rmdefault}{\mddefault}{\updefault}{\color[rgb]{0,0,0}$C_{A,\{m_1\},1}$}%
}}}}
\put(2701,-1321){\makebox(0,0)[lb]{\smash{{\SetFigFont{10}{12.0}{\rmdefault}{\mddefault}{\updefault}{\color[rgb]{0,0,0}$C_{B,\emptyset,1}$}%
}}}}
\put(4501,-1321){\makebox(0,0)[lb]{\smash{{\SetFigFont{10}{12.0}{\rmdefault}{\mddefault}{\updefault}{\color[rgb]{0,0,0}$C_{B,\{m_2\},1}$}%
}}}}
\put(4501,-1141){\makebox(0,0)[lb]{\smash{{\SetFigFont{10}{12.0}{\rmdefault}{\mddefault}{\updefault}{\color[rgb]{0,0,0}$C_{A,\{m_1\},2}$}%
}}}}
\put(4501,434){\makebox(0,0)[lb]{\smash{{\SetFigFont{10}{12.0}{\rmdefault}{\mddefault}{\updefault}{\color[rgb]{0,0,0}$\slide(B)$}%
}}}}
\put(4591,-2941){\makebox(0,0)[lb]{\smash{{\SetFigFont{10}{12.0}{\rmdefault}{\mddefault}{\updefault}{\color[rgb]{0,0,0}$C_{A,\{m_1,m_3\},2}$}%
}}}}
\put(4591,-3121){\makebox(0,0)[lb]{\smash{{\SetFigFont{10}{12.0}{\rmdefault}{\mddefault}{\updefault}{\color[rgb]{0,0,0}$C_{B,\{m_2\},0}$}%
}}}}
\put(2701,-2941){\makebox(0,0)[lb]{\smash{{\SetFigFont{10}{12.0}{\rmdefault}{\mddefault}{\updefault}{\color[rgb]{0,0,0}$C_{A,\{m_1,m_3\},1}$}%
}}}}
\put(2701,-3121){\makebox(0,0)[lb]{\smash{{\SetFigFont{10}{12.0}{\rmdefault}{\mddefault}{\updefault}{\color[rgb]{0,0,0}$C_{B,\emptyset,0}$}%
}}}}
\end{picture}%

%% file: conclusion.tex
\section{Conclusion}
\label{sec:conclusion}

In this paper, we developed a suitable approach based on biological
data taking into account durations to identify new constraints on
parameters piloting the dynamics of the model. We proved the
usefulness of our hybrid Hoare logic by determining the
constraints on celerities of a small interaction graph
between the \textit{lacI} repressor and the \textit{Ntr} system. 
Indeed, our backward strategy leads to a constraint for every
celerity encountered along the observed cyclic behaviour.

To strengthen our formalism, we now have to prove
the soundness and correctness of our weakest precondition calculus.
One of the drawbacks of our Hoare logic resides in the size of the
obtained precondition. It becomes mandatory to simplify on the fly all
intermediate formulas. To go further, it would be useful to give the
obtained weakest precondition to a constraint solver in order to
generate one or several solutions. Thereafter one can automate the
identification of celerities and simulations to compare the simulated
traces with experimental biological data. This global process
has already been performed manually on a model of
the cell cycle in mammals~\cite{behaegel2016hybrid}.

Beyond the sequential path studied into this paper, our approach may
be extended to the conditional branching instruction
($if\ [\ ]\ then\ [\ ]\ else\ [\ ]$), or the iteration instruction
($while\ [\ ]\ do\ [\ ]$) as it was already made in the discrete
case~\cite{inviteCMSB2015}. This should lead to an expressive framework in
which durations between two experimental measures, as well as
complex behaviours, are well taken into consideration.

%% file: formulas.tex
\section{Sub-properties of the Weakest Precondition Calculus}
\label{ap:subformulas}

In this appendix, we detail the formal content of the sub-properties
$\Phi_v^+(\Delta t)$, $\Phi_v^-(\Delta t)$, $\F_v(\Delta t)$,
$\W^+_v$, $\W^-_v$, $\A(\Delta t, assert)$ and $\Rep_v$
informally presented in \pref{ssec:wp}.

It has to be noted that all of these properties depend on the
indices $i$ and $f$ used in \pref{def:wp},
although for readability issues we did not mention them
on the names of each sub-property.
Furthermore, for a given index $i$,
we call by convention $\pii{u}{i}$ (\resp $\pii{u}{i}'$)
the fractional part of the exiting (\resp entering) state
inside the discrete state $i$.
Finally, we recall that $\Phi_u^{\omega}$
is true iff the resources of $u$ in the current state are exactly $\omega$,
as formally expressed in \pref{def:resources}.

\subsection{Discrete Transition to the Next Discrete State}
\label{ssec:phi}

\newcommand{\forphi}[3]{%
  \Phi_v^{#1}(\Delta t) &\equiv (\pii{v}{i} = #2) \\ &\quad \wedge
  ~\bigwedge_{\mathclap{\substack{\omega \subset \pred(v)\\n \in \Nsegm{0}{b_v}}}}
  \Big( \big(\Phi_v^\omega\wedge(\eta_v=n)\big)
  \Rightarrow (C_{v,\omega,n} #3 0) \wedge (\pii{v}{i}' = \pii{v}{i} - C_{v, \omega, n} \cdot \Delta t) \Big)
}

For all component $v \in V$,
$\Phi_v^+(\Delta t)$ (\resp $\Phi_v^-(\Delta t)$) describes the conditions in which
$v$ increases (\resp decreases) its discrete expression level:
its celerity in the current state must be positive (\resp negative)
and its fractional part only depends on $\Delta t$
in the way given at the very end of \pref{sec:framework}.
\begin{align*}
  \forphi{+}{1}{>} \enspace, \\
  \forphi{-}{0}{<} \enspace.
\end{align*}

\newcommand{\forsubfirst}[3]{%
  \Big( \bigwedge_{\mathclap{\substack{\omega \subset \pred(u)\\n \in \Nsegm{0}{b_u}}}}
  \big((\eta_u = n) \wedge \Phi_u^\omega \wedge C_{u, \omega, n} #1 0 \wedge
  \pi'_{u,i} #2 \pi_{u,i} - C_{u, \omega, n} \cdot \Delta t \big)
  \Rightarrow \W^{#3}_u \Big)
}

\newcommand{\forew}[3]{%
  \EW^{#1}_u &\equiv
    (\eta_u = #2) \wedge \bigwedge_{\mathclap{\omega \subset \pred(u)}}
    \left(\Phi_u^\omega \Rightarrow
    C_{u, \omega, #2} #3 0 \right)
}
\newcommand{\foriw}[4]{%
  \IW^{#1}_u &\equiv (\eta_u #2) \wedge
    \bigwedge_{\mathclap{\substack{\omega, \omega' \subset \pred(u)\\n \in \Nsegm{0}{b_u}}}} \Big( \big((\eta_u = n) \wedge (m = n #1 1) 
    \wedge \Phi_u^\omega \wedge \Phi_{u #1}^{\omega'} \big) \\
  &\qquad\qquad\qquad\qquad
    \Rightarrow C_{u, \omega, n} #3 0 \wedge C_{u, \omega', m} #4 0 \Big)
}
\newcommand{\forphibis}[2]{%
  \Phi_{u #1}^{\omega'} &\equiv (\eta_u #2) \wedge
    \bigwedge_{\mathclap{n \in \Nsegm{0}{b_u}}} \big((\eta_u = n) \Rightarrow
    \Phi_u^{\omega'} [\eta_u \backslash \eta_u #1 1] \big)
}

\subsection{Internal and External Walls}
\label{ssec:walls}

For all component $u \in V$,
$\W^+_u$ (\resp $\W^-_u$) states that there is a wall
preventing $u$ to increase (\resp decrease) its qualitative state.
This wall can either be
an external wall $\EW^+_u$ (\resp $\EW^-_u$)
or an internal wall $\IW^+_u$ (\resp $\IW^-_u$).
Furthermore, $\Phi_{u+}^{\omega'}$ (\resp $\Phi_{u-}^{\omega'}$),
which is required in these sub-formulas,
is true if and only if the set of resources of $u$ is exactly $\omega'$
in the state where $u$ is increased (\resp decreased) by 1.
\[\W^{+}_u \equiv \BW^{+}_u \vee \EB^{+}_u
  \text{\qquad and \qquad}
  \W^{-}_u \equiv \BW^{-}_u \vee \EB^{-}_u\]
where:
\begin{align*}
  \forew{+}{b_u}{>} \enspace, \\
  \forew{-}{0}{<} \enspace,
\end{align*}
\begin{align*}
  \foriw{+}{< b_u}{>}{<} \enspace, \\
  \foriw{-}{> 0}{<}{>} \enspace,
\end{align*}
\begin{align*}
  \forphibis{+}{< b_u} \enspace, \\
  \forphibis{-}{> 0} \enspace.
\end{align*}

\newcommand{\forslide}[3]{%
  \Sl^{#1}_{u, \omega_u, n_u}(\Delta t) &\equiv \pii{u}{i} = #2 \wedge
    \big( C_{u,\omega,n} #3 0
    \Rightarrow \pii{u}{i}' #3 \pii{u}{i} - C_{u,\omega,n} \cdot \Delta t \big)
}

\subsection{Knocking Variable Able to Perform a Discrete Transition}
\label{ssec:first}

For all component $v \in V$,
$\F_v(\Delta t)$ states that all components other than $v$ must either
reach their border after $v$, or face an internal or external wall.
In other words, $v$ belongs to the set of components which can first
change its qualitative state.
This is required to enforce the meaning of an instruction $v+$ or $v-$,
which means that no other component changes its qualitative state before~$v$.
\begin{align*}
  &\F_v(\Delta t) \equiv \bigwedge_{\mathclap{u \in V \setminus \{ v \}}} \\
    &\quad \quad \forsubfirst{>}{>}{+} \\
    &\quad \wedge \forsubfirst{<}{<}{-} \enspace.
\end{align*}

\subsection{Hybrid Assertions}
\label{ssec:assert}

The sub-property $\A(\Delta t, a)$ allows one to translate all assertion symbols
given about the continuous transition related to the instruction
(celerities and slides) into a property:
\[
  \A(\Delta t, a) \equiv \bigwedge_{\mathclap{\mbox{\scriptsize$\begin{matrix*}[r]u \in V \\n_u \in \Nsegm{0}{b_u}\\\omega_u \in \pred(u)\end{matrix*}$}~~~~~~~~}}~~
    \Big(\Big(\bigwedge_{\mathclap{~~~~~\mbox{\scriptsize$\begin{matrix*}[l]
      v \in V \\
      n_v \in \Nsegm{0}{b_v} \\
      \omega_v \in \pred(v)
    \end{matrix*}$}}}
    \Big((\eta_v = n_v) \wedge \Phi_v^{\omega_v}
    \Rightarrow a[C_v \backslash C_{v,\omega_v,n_v}]\Big)\Big)
    \!\!\left[\begin{matrix}
      \slide(u) \backslash \Sl_{u, \omega_u, n_u} \\
      \slide^+(u) \backslash \Sl^+_{u, \omega_u, n_u} \\
      \slide^-(u) \backslash \Sl^-_{u, \omega_u, n_u} \\
    \end{matrix}\right]\!\!\Big)
\]
where $A$ is the assert part of the instruction, that is,
$P = (\Delta t, A, v+)$ or $P = (\Delta t, A, v-)$, and, for all variable $u \in V$:
\begin{align*}
  \forslide{+}{1}{>} \enspace, \\
  \forslide{-}{0}{<} \enspace, \\
  \Sl_{u, \omega_u, n_u}(\Delta t) &\equiv \Sl^{+}_{u, \omega_u, n_u}(\Delta t) \vee \Sl^{-}_{u, \omega_u, n_u}(\Delta t) \enspace.
\end{align*}

\subsection{Junctions Between Discrete States}
\label{ssec:junc}

For all component $v \in V$,
and for a discrete transition happening on component $v$ between
an initial and a final state corresponding to the indices $i$ and $f$,
$\Rep_v$ establishes a junction between the fractional parts of these states.
It states that the fractional part of $v$ switches from $1$ to $0$ for an increase,
or from $0$ to $1$ for a decrease, and all other fractional parts are unchanged:
\[\Rep_v \equiv (\pii{v}{f} = 1 - \pii{v}{i}')
  \wedge \bigwedge_{\mathclap{u \in V \setminus \{ v \}}} (\pii{u}{f} = \pii{u}{i}') \enspace.
\]

The phenomenon described by $\Rep_v$ can be easily observed on \pref{fig:pathExample}
on the discrete transition in the centre:
all fractional parts are left the same,
except for the variable performing the transition.

%% file: appendix.tex
\section{Computation Steps of the \textit{lacI} Repressor Example}
\label{ap:steps}

\subsection{Computation of $D_1$ and $H_1$}
\label{ap:step1}

We detail here the computations that lead to the results summed up
in \pref{ssec:step1example}. For better readability, we simplified 
the numerous conjunctions produced
by removing the terms that were trivially true,
mainly due to implications with a false premise
because of terms such as
$\Phi_{v}^{\omega}$, $\Phi_{v+}^{\omega}$, $\Phi_{v-}^{\omega}$,
 $m = int$, $\eta_{var} = int$ and 
$C_{v,\omega,n}\ \square\ cste$.
In addition, the notation $[\ldots]$ depicts a sub-property
which is not expanded because it is in conjunction
with another trivially false property.
This step concerns the first step of the Hoare triple previously 
presented, which corresponds to:
\[
  \prop{D_1}{H_1}
  \patha{T_1}{\top}{A+}
  \prop{D_0 \equiv (\eta_A=2 \wedge \eta_B=0)}{H_0 \equiv \top}
    \enspace.
\]

At this point, \pref{def:wp} permits to find the
weakest precondition, which is given by:
\begin{align}
  D_1 & \equiv D_0 [\eta_A \backslash \eta_A+1] \equiv (\eta_A+1=2) \wedge (\eta_B=0) \equiv 
    (\eta_A=1) \wedge (\eta_B=0) \enspace, \\
  H_1 & \equiv H_0 \wedge \Phi_A^+(T_1) \wedge \F_A(T_1)
    \wedge \neg \W^+_A \wedge \A(T_1) \wedge \Rep_A \enspace.
\end{align}

Firstly, we calculated each element of the hybrid part $H_1$. The $\EB$ and 
$\BW$ formulas (formally given in \pref{ssec:walls}) are computed for 
both variables. These formulas will be used thereafter in
the sub-formulas $\F_A(T_1)$ and $\neg \W^+_A$.

\begin{align*}
  \BW^+_A &\equiv \left( 
  \begin{array}{r}
    (\eta_A < b_A) \wedge \big( (\eta_A = 1) \wedge (m=2) 
  \wedge \Phi_A^{m_1,m_3} \wedge \Phi_{A+}^{m_1,m_3} \\
    \Rightarrow C_{A,\{m_1,m_3\},1} > 0 \wedge 
  C_{A,\{m_1,m_3\},2} < 0\big)
  \end{array} \right) 
  \equiv \bot\\
  \BW^-_B &\equiv (\eta_B > 0) \wedge [\ldots] \equiv \bot \\
  \BW^+_B &\equiv \left(
  \begin{array}{r}
    (\eta_B < b_B) \wedge \big( (\eta_B = 0) \wedge( m=1) 
    \wedge \Phi_B^\emptyset \wedge \Phi_{B+}^\emptyset \\
    \Rightarrow C_{B,\emptyset,0} > 0 \wedge C_{B,\emptyset,1} 
    < 0 \big)
  \end{array} \right)
  \equiv \bot\\
  \EB^+_A &\equiv (\eta_A = b_A) \wedge [\ldots] \equiv \bot \\
  \EB^-_B &\equiv (\eta_B = 0) \wedge \big( \Phi_B^\emptyset \Rightarrow 
  C_{B,\emptyset,0} <0 \big) \equiv C_{B,\emptyset,0} <0\\
  \EB^+_B &\equiv (\eta_B = b_B) \wedge [\ldots] \equiv \bot
\label{eq:ExampleComputationWall}
\end{align*}

Formulas $\BW^+_A$ and $\BW^+_B$ are false because all 
the conditions in the premises are true but their conclusion are false.
Indeed, celerities having the same resources despite a different
qualitative level should have the same nonzero sign
(see \pref{def:network}).
Thus, $(C_{A,\{m_1,m_3\},1} > 0 \wedge C_{A,\{m_1,m_3\},2} < 0)$
as well as
$(C_{B,\emptyset,0} > 0 \wedge C_{B,\emptyset,1} < 0)$
can be reduced to false.

Moreover, formulas $(\eta_A = b_A)$, $(\eta_B = b_B)$ and $(\eta_B > 0)$ are 
false because $(\eta_A = 1 < b_A)$ and $(\eta_B = 0)$ in the 
considered state which is $(\eta_A=1) \wedge (\eta_B=0)$,
therefore $\EB^+_A$, $\EB^+_B$ and $\BW^-_B$ are
false. Lastly, the computation of $\EB^-_B$ gives the constraint 
$C_{B,\emptyset,0} < 0$.

Using the formulas of \pref{ssec:walls} and the previous results, we deduce that 
$\neg \W^+_A \equiv \neg \BW^+_A \wedge \neg \EB^+_A \equiv \top$,  
$\W^+_B \equiv \bot$ and $\W^-_B \equiv C_{B,\emptyset,0} <0$.

\begin{itemize}
  \item Computation of $\F_A(T_1)$: Because $(\eta_B = 0)$ and $\Phi_B^\emptyset$
  are $true$, we have
    \begin{equation}
    \begin{aligned}
      \F_A(T_1) &\equiv \big( (C_{B,\emptyset,0}>0) \wedge (\pii{B}{1}' > 
      \pii{B}{1} - C_{B,\emptyset,0} \cdot T_1) \big) \Rightarrow \W^+_B\\
      &\wedge \big( (C_{B,\emptyset,0}<0) \wedge (\pii{B}{1}' < \pii{B}{1} 
      - C_{B,\emptyset,0} \cdot T_1) \big) \Rightarrow \W^-_B\\
      &\equiv \neg (C_{B,\emptyset,0} > 0) \vee \neg (\pii{B}{1}' > \pii{B}{1} - 
    C_{B,\emptyset, 0} \cdot T_1) \enspace.
    \end{aligned}
    \label{eq:ExampleComputationFirst}
    \end{equation}

  \item Computation of $\Phi_A^+(T_1)$:
    \begin{equation}\label{eq:ExampleComputationNextState}
    \begin{aligned}
      \Phi_A^+(T_1) &\equiv (\pii{A}{1} = 1) \wedge \Big( \Phi_A^{\{m_1,m_3\}} \wedge 
      (\eta_A = 1) \\
      &\qquad \Rightarrow (C_{A,\{m_1,m_3\},1} > 0) \wedge (\pii{A}{1}' 
      = \pii{A}{1} - C_{A,\{m_1,m_3\},1} \cdot T_1) \Big ) \enspace.
    \end{aligned}
    \end{equation}
    Because $\Phi_A^{\{m_1,m_3\}} \equiv \top$ and $(\eta_A = 1) \equiv \top$, 
    we deduce:
    \[\Phi_A^+(T_1) \equiv (\pii{A}{1} = 1) \wedge (C_{A,\{m_1,m_3\},1} > 0) 
    \wedge (\pii{A}{1}' = \pii{A}{1} - C_{A,\{m_1,m_3\},1} \cdot T_1) \enspace.\]

  \item Computation of $\Rep_A$:
    \begin{equation}\label{eq:ExampleComputationJunction}
      \Rep_A \equiv (\pii{B}{1} = \pii{B}{0}') \wedge (\pii{A}{1} = 1-\pii{A}{0}') \enspace.
    \end{equation}
    Equation~\eqref{eq:ExampleComputationJunction} links the fractional 
    part of $B$ in the current discrete state with the next one
    of the execution path. The 
    fractional part of $A$ is also linked to the next state,
    whose value is know due to the formula of \pref{ssec:phi}.

  \item For this particular path, $\A(T_1) \equiv \top$.
\end{itemize}

We can deduce the value of the hybrid part $H_1$:
\begin{equation}\label{eq:AppendixExampleResult}
\begin{aligned}
  H_1 &\equiv \Big( \neg 
  (C_{B,\emptyset,0} > 0) \vee \neg (\pii{B}{1}' > \pii{B}{0}' - 
  C_{B,\emptyset,0} \cdot T_1) \Big)\\
  &\wedge (C_{A,\{m_1,m_3\},1} > 0) \wedge (\pii{A}{1}' = 1 - 
  C_{A,\{m_1,m_3\},1} \cdot T_1) \wedge (\pii{A}{0}' = 0) \enspace.
\end{aligned}
\end{equation}

Therefore, we obtained constraints for each celerity as well as each fractional 
part of this current state.

\subsection{Computation of $D_2$ and $H_2$}
\label{ap:step2}

\[
  \prop{D_2}{H_2}
  \patha{T_2}{\top}{B-}
  \prop{D_1 \equiv (\eta_A=1 \wedge \eta_B=0)}{H_1}
\]

\begin{align}
  D_2 & \equiv D_1[\eta_B \backslash \eta_B-1] \equiv (\eta_A=1) \wedge (\eta_B-1=0) \equiv (\eta_A=1) \wedge (\eta_B=1) \\
  H_2 & \equiv H_1 \wedge \F_B(T_2) \wedge \Phi_B^-(T_2) \wedge \neg \W^-_B \wedge \A(T_2) \wedge \Rep_B
\end{align}

We calculated the formulas necessary to define $\F_B(T_2)$ and $\neg \W^-_B$:

\begin{align*}
  \BW^-_A &\equiv (\eta_A > 0) \wedge \big( (\eta_A = 1) \wedge (m=0) \wedge \Phi_A^{m_1} \wedge \Phi_{A-}^\emptyset\\
  &\qquad\qquad\qquad \Rightarrow C_{A,\{m_1\},1} < 0 \wedge C_{A,\emptyset,0} > 0 \big) \\
  \BW^+_A &\equiv (\eta_A < b_A) \wedge \big( (\eta_A = 1) \wedge (m=2) \wedge \Phi_A^{m_1} \wedge \Phi_{A+}^{m_1}\\
  &\qquad\qquad\qquad \Rightarrow C_{A,\{m_1\},1} > 0 \wedge C_{A,\{m_1\},2} < 0 \big) \\
  &\equiv \bot\\
  \BW^-_B &\equiv (\eta_B > 0) \wedge \big( (\eta_B = 1) \wedge (m=0) \wedge \Phi_B^\emptyset \wedge \Phi_{B-}^\emptyset \\
  &\qquad\qquad\qquad\Rightarrow C_{B,\emptyset,1} < 0 \wedge C_{B,\emptyset,0} > 0\big)\\
  &\equiv \bot\\
  \EB^-_A &\equiv (\eta_A = 0) \wedge [\ldots] \equiv \bot\\
  \EB^+_A &\equiv (\eta_A = b_A) \wedge [\ldots] \equiv \bot \\
  \EB^-_B &\equiv (\eta_B = 0) \wedge [\ldots] \equiv \bot \\
\end{align*}

With these results, we deduce that $\neg \W^-_B \equiv \neg \BW^-_B \wedge \neg \EB^-_B \equiv \top$.

\begin{itemize}
  \item Computation of $\F_B(T_2)$:
    \begin{equation}\label{}
      \begin{aligned}
        \F_B(T_2) &\equiv \big( (\eta_A = 1) \wedge \Phi_A^{m_1} \wedge (C_{A,\{m_1\},1}>0) \wedge (\pii{A}{2}' > \pii{A}{2} - C_{A,\{m_1\},1} \cdot T_2) \big) \\
        &\qquad\qquad \Rightarrow \W^+_A\\
        \wedge& \big( (\eta_A = 1) \wedge \Phi_A^{m_1} \wedge (C_{A,\{m_1\},1}<0) \wedge (\pii{A}{2}' < \pii{A}{2} - C_{A,\{m_1\},1} \cdot T_2) \big) \\
        &\qquad\qquad \Rightarrow \W^-_A
      \end{aligned}      
    \end{equation}
    We deduce that $\F_B(T_2) \equiv \big( \neg (C_{A,\{m_1\},1}>0) \vee \neg (\pii{A}{2}' > \pii{A}{2} - C_{A,\{m_1\},1} \cdot T_2) \big) \wedge \Big((C_{A,\emptyset,0} > 0) \vee \neg (C_{A,\{m_1\},1}<0) \vee \neg (\pii{A}{2}' < \pii{A}{2} - C_{A,\{m_1\},1} \cdot T_2) \Big)$.

  \item Computation of $\Phi_B^-(T_2)$:
    \begin{equation}\label{}
    \begin{aligned}
      \Phi_B^-(T_2) &\equiv (\pii{B}{2} = 0) \wedge \big( \Phi_B^\emptyset \wedge (\eta_B = 1) \\
      &\qquad\qquad \Rightarrow (C_{B,\emptyset,1} < 0) \wedge (\pii{B}{2}' = \pii{B}{2} - C_{B,\emptyset,1} \cdot T_2) \big)\\
    \end{aligned}
    \end{equation}
  Because $\Phi_B^\emptyset \equiv \top$ and $(\eta_B = 1) \equiv \top$, we deduce:\\
  $\Phi_B^-(T_2) \equiv (\pii{B}{2} = 0) \wedge (C_{B,\emptyset,1} < 0) \wedge (\pii{B}{2}' = \pii{B}{2} - C_{B,\emptyset,1} \cdot T_2)$

  \item Computation of $\Rep_B$:
  \begin{equation}\label{}
    \Rep_B \equiv (\pii{A}{2} = \pii{A}{1}') \wedge (\pii{B}{2} = 1-\pii{B}{1}')
  \end{equation}

  \item For this particular path, $\A(T_2) \equiv \top$.
\end{itemize}

We can deduce the value of the hybrid part $H_2$:

\begin{align}
\begin{split}
  H_2 &\equiv H_1 \wedge \big( \neg (C_{A,\{m_1\},1}>0) \vee \neg (\pii{A}{2}' > \pii{A}{2} - C_{A,\{m_1\},1} \cdot T_2) \big) \\
  &\wedge \Big((C_{A,\emptyset,0} > 0) \vee \neg (C_{A,\{m_1\},1}<0) \vee \neg (\pii{A}{2}' < \pii{A}{2} - C_{A,\{m_1\},1} \cdot T_2) \Big) \\
  &\wedge (\pii{B}{2} = 0) \wedge (C_{B,\emptyset,1} < 0) \wedge (\pii{B}{2}' = \pii{B}{2} - C_{B,\emptyset,1} \cdot T_2) \\  
  &\wedge (\pii{A}{2} = \pii{A}{1}') \wedge (\pii{B}{1}' = 1) \\
\end{split} \\
\begin{split}
  H_2 &\equiv \Big( \neg (C_{B,\emptyset,0} > 0) \vee \neg (1 > \pii{B}{0}' - C_{B,\emptyset,0} \cdot T_1) \Big)\\
  &\wedge (C_{A,\{m_1,m_3\},1} > 0) \wedge (\pii{A}{1}' = 1 - C_{A,\{m_1,m_3\},1} \cdot T_1) \wedge (\pii{A}{0}' = 0)\\
  &\wedge \big( \neg (C_{A,\{m_1\},1}>0) \vee \neg (\pii{A}{2}' > \pii{A}{1}' - C_{A,\{m_1\},1} \cdot T_2) \big) \\
  &\wedge \Big((C_{A,\emptyset,0} > 0) \vee \neg (C_{A,\{m_1\},1}<0) \vee \neg (\pii{A}{2}' < \pii{A}{1}' - C_{A,\{m_1\},1} \cdot T_2) \Big) \\
  &\wedge (C_{B,\emptyset,1} < 0) \wedge (\pii{B}{2}' = 0 - C_{B,\emptyset,1} \cdot T_2) \\
\end{split}
\end{align}

\subsection{Computation of $D_3$ and $H_3$}
\label{ap:step3}

\[
  \prop{D_3}{H_3}
  \patha{T_3}{\slide^+(B)}{A-}
  \prop{D_2 \equiv (\eta_A=1 \wedge \eta_B=1)}{H_2}
\]

\begin{align}
  D_3 & \equiv D_2[\eta_A \backslash \eta_A-1] \equiv (\eta_A-1=1) \wedge (\eta_B=1) \equiv (\eta_A=2) \wedge (\eta_B=1) \\
  H_3 & \equiv H_2 \wedge \F_A(T_3) \wedge \Phi_A^-(T_3) \wedge \neg \W^-_A \wedge \A(T_3) \wedge \Rep_A
\end{align}

We calculated the formulas necessary to define $\F_A(T_3)$ and $\neg \W^-_A$:

\begin{align*}
  \BW^-_A &\equiv (\eta_A > 0) \wedge \big( (\eta_A = 2) \wedge (m=1) \wedge \Phi_A^{m_1} \wedge \Phi_{A-}^{m_1} \\
  &\qquad\qquad\qquad \Rightarrow C_{A,\{m_1\},2} < 0 \wedge C_{A,\{m_1\},1} > 0\big)\\
  &\equiv \bot\\
  \BW^-_B &\equiv (\eta_B > 0) \wedge \big( \eta_B = 1 \wedge m=0 \wedge \Phi_B^{m_2} \wedge \Phi_{B-}^{m_2}\\
  &\qquad\qquad\qquad \Rightarrow C_{B,\{m_2\},1} < 0 \wedge C_{B,\{m_2\},0} > 0 \big) \\
  &\equiv \bot\\
  \BW^+_B &\equiv (\eta_B < b_B) \wedge [\ldots] \equiv \bot \\
  \EB^-_A &\equiv (\eta_A = 0) \wedge [\ldots] \equiv \bot \\
  \EB^-_B &\equiv (\eta_B = 0) \wedge [\ldots] \equiv \bot\\
  \EB^+_B &\equiv (\eta_B = b_B) \wedge (\Phi_B^{m_2} \Rightarrow C_{B,\{m_2\},1 > 0})
\end{align*}

With these results, we deduce that $\neg \W^-_A \equiv \neg \BW^-_A \wedge \neg \EB^-_A \equiv \top$.

\begin{itemize}
  \item Computation of $\F_A(T_3)$:
    \begin{equation}\label{}
    \begin{aligned}
      \F_A(T_3) &\equiv \big( (\eta_B = 1) \wedge \Phi_B^{m_2} \wedge (C_{B,\{m_2\},1}>0) \wedge (\pii{B}{3}' > \pii{B}{3} - C_{B,\{m_2\},1} \cdot T_3) \big)\\
      &\qquad\qquad \Rightarrow \W^+_B\\
      \wedge& \big( (\eta_B = 1) \wedge \Phi_B^{m_2} \wedge (C_{B,\{m_2\},1}<0) \wedge (\pii{B}{3}' < \pii{B}{3} - C_{B,\{m_2\},1} \cdot T_3) \big)\\
      &\qquad\qquad \Rightarrow \W^-_B
    \end{aligned}
    \end{equation}
  We deduce that $\F_A(T_3) \equiv \big( \neg (C_{B,\{m_2\},1}<0) \vee \neg (\pii{B}{3}' < \pii{B}{3} - C_{B,\{m_2\},1} \cdot T_3) \big)$.

  \item Computation of $\Phi_A^-(T_3)$:
    \begin{equation}\label{}
    \begin{aligned}
      \Phi_A^-(T_3) &\equiv (\pii{A}{3} = 0) \wedge \big( \Phi_A^{m_1} \wedge (\eta_A = 2) \\
      &\qquad\qquad \Rightarrow (C_{A,\{m_1\},2} < 0) \wedge (\pii{A}{3}' = \pii{A}{3} - C_{A,\{m_1\},2} \cdot T_3) \big)\\
    \end{aligned}
    \end{equation}
  Because $\Phi_A^{m_1} \equiv \top$ and $(\eta_A = 2) \equiv \top$, we deduce:\\
  $\Phi_A^-(T_3) \equiv (\pii{A}{3} = 0) \wedge \big( C_{A,\{m_1\},2} < 0 \wedge (\pii{A}{3}' = \pii{A}{3} - C_{A,\{m_1\},2} \cdot T_3) \big) \\$

  \item Computation of $\Rep_A$:
    \begin{equation}\label{}
      \Rep_A \equiv (\pii{B}{3} = \pii{B}{2}') \wedge (\pii{A}{3} = 1 - \pii{A}{2}')
    \end{equation}

  \item Computation of $\A(T_3)$:
    \begin{equation}\label{}
    \begin{aligned}
      \A(T_3) &\equiv \Sl^+_B\\
      &\equiv (\pii{B}{3} = 1) \wedge \big( (\eta_B = 1) \wedge \Phi_B^{m_2} \\
      &\qquad\qquad \Rightarrow (C_{B,\{m_2\},1} > 0 \Rightarrow \pii{B}{3}' > \pii{B}{3} - C_{B,\{m_2\},1} \cdot T_3) \big)\\
      &\equiv (\pii{B}{3} = 1) \\
      &\wedge \Big( \neg (C_{B,\{m_2\},1} > 0) \vee (\pii{B}{3}' > \pii{B}{3} - C_{B,\{m_2\},1} \cdot T_3) \Big) \\
    \end{aligned}
    \end{equation}
\end{itemize}

We can deduce the value of the hybrid part $H_3$:

\begin{align}
\begin{split}
  H_3 &\equiv H_2 \wedge \big( \neg (C_{B,\{m_2\},1}<0) \vee \neg (\pii{B}{3}' < \pii{B}{3} - C_{B,\{m_2\},1} \cdot T_3) \big)\\
  &\wedge (\pii{A}{3} = 0) \wedge \big( C_{A,\{m_1\},2} < 0 \wedge (\pii{A}{3}' = \pii{A}{3} - C_{A,\{m_1\},2} \cdot T_3) \big) \\
  &\wedge (\pii{B}{3} = 1) \wedge \Big( \neg (C_{B,\{m_2\},1} > 0) \vee (\pii{B}{3}' > \pii{B}{3} - C_{B,\{m_2\},1} \cdot T_3) \Big) \\
  &\wedge (\pii{B}{3} = \pii{B}{2}') \wedge (\pii{A}{2}' = 1)\\
\end{split} \\
\begin{split}
  H_3 &\equiv \Big( \neg (C_{B,\emptyset,0} > 0) \vee \neg (1 > \pii{B}{0}' - C_{B,\emptyset,0} \cdot T_1) \Big)\\
  &\wedge (C_{A,\{m_1,m_3\},1} > 0) \wedge (\pii{A}{1}' = 1 - C_{A,\{m_1,m_3\},1} \cdot T_1) \wedge (\pii{A}{0}' = 0)\\
  &\wedge \big( \neg (C_{A,\{m_1\},1}>0) \vee \neg (1 > \pii{A}{1}' - C_{A,\{m_1\},1} \cdot T_2) \big) \\
  &\wedge \Big((C_{A,\emptyset,0} > 0) \vee \neg (C_{A,\{m_1\},1}<0) \vee \neg (1 < \pii{A}{1}' - C_{A,\{m_1\},1} \cdot T_2) \Big) \\
  &\wedge (C_{B,\emptyset,1} < 0) \wedge (1 = 0 - C_{B,\emptyset,1} \cdot T_2) \\
  &\wedge \big( \neg (C_{B,\{m_2\},1}<0) \vee \neg (\pii{B}{3}' < 1 - C_{B,\{m_2\},1} \cdot T_3) \big)\\
  &\wedge \big( C_{A,\{m_1\},2} < 0 \wedge (\pii{A}{3}' = 0 - C_{A,\{m_1\},2} \cdot T_3) \big) \\
  &\wedge \Big( \neg (C_{B,\{m_2\},1} > 0) \vee (\pii{B}{3}' > 1 - C_{B,\{m_2\},1} \cdot T_3) \Big)\\
\end{split}
\end{align}

\subsection{Computation of $D_4$ and $H_4$}
\label{ap:step4}

\[
  \prop{D_4}{H_4}
  \patha{T_4}{\top}{B+}
  \prop{D_3 \equiv (\eta_A=2 \wedge \eta_B=1)}{H_3}
\]

\begin{align}
  D_4 & \equiv D_3[\eta_B \backslash \eta_B+1] \equiv (\eta_A=2) \wedge (\eta_B+1=1) \equiv (\eta_A=2) \wedge (\eta_B=0) \\
  H_4 & \equiv H_3 \wedge \F_B(T_4) \wedge \Phi_B^+(T_4) \wedge \neg \W^+_B \wedge \A(T_4) \wedge \Rep_B
\end{align}

We calculated the formulas necessary to define $\F_B(T_4)$ and $\neg \W^+_B$:

\begin{align*}
  \BW^-_A &\equiv (\eta_A > 0) \wedge \big( (\eta_A = 2) \wedge (m=1) \wedge \Phi_A^{m_1,m_3} \wedge \Phi_{A-}^{m_1,m_3} \\
  &\qquad\qquad\qquad \Rightarrow (C_{A,\{m_1,m_3\},2} < 0) \wedge (C_{A,\{m_1,m_3\},1} > 0) \big) \\
  &\equiv \bot\\
  \BW^+_A &\equiv (\eta_A < b_A) \wedge [\ldots] \equiv \bot \\
  \BW^+_B &\equiv (\eta_B < b_B) \wedge \big( (\eta_B = 0) \wedge (m=1) \wedge \Phi_B^{m_2} \wedge \Phi_{B+}^{m_2} \\
  &\qquad\qquad\qquad \Rightarrow C_{B,\{m_2\},0} > 0 \wedge C_{B,\{m_2\},1} < 0\big)\\
  &\equiv \bot\\
  \EB^-_A &\equiv (\eta_A = 0) \wedge [\ldots] \equiv \bot\\
  \EB^+_A &\equiv (\eta_A = b_A) \wedge (\Phi_A^{m_1,m_3} \Rightarrow C_{A,\{m_1,m_3\},2} > 0)\\
  \EB^+_B &\equiv (\eta_B = b_B) \wedge [\ldots] \equiv \bot
\end{align*}

With these results, we deduce that $\neg \W^+_B \equiv \neg \BW^+_B \wedge \neg \EB^+_B \equiv \top$.

\begin{itemize}
  \item Computation of $\F_B(T_4)$:
    \begin{equation}\label{}
    \begin{aligned}
      \F_B(T_4) &\equiv \big( (\eta_A = 2) \wedge \Phi_A^{m_1,m_3} \wedge (C_{A,\{m_1,m_3\},2}>0) \\
      &\wedge (\pii{A}{4}' > \pii{A}{4} - C_{A,\{m_1,m_3\},2} \cdot T_4) \big) \Rightarrow \W^+_A\\
      &\wedge \big( (\eta_A = 2) \wedge \Phi_A^{m_1,m_3} \wedge (C_{A,\{m_1,m_3\},2}<0) \\
      &\wedge (\pii{A}{4}' < \pii{A}{4} - C_{A,\{m_1,m_3\},2} \cdot T_4) \big) \Rightarrow \W^-_A
    \end{aligned}
    \end{equation}

  We deduce that $\F_B(T_4) \equiv \big( \neg (C_{A,\{m_1,m_3\},2}<0) \vee \neg (\pii{A}{4}' < \pii{A}{4} - C_{A,\{m_1,m_3\},2} \cdot T_4) \big)$.

  \item Computation of $\Phi_B^+(T_4)$:
    \begin{equation}\label{}
    \begin{aligned}
      \Phi_B^+(T_4) &\equiv (\pii{B}{4} = 1) \wedge \big( \Phi_B^{m_2} \wedge (\eta_B = 0) \\
      &\qquad\qquad \Rightarrow (C_{B,\{m_2\},0} > 0) \wedge (\pii{B}{4}' = \pii{B}{4} - C_{B,\{m_2\},0} \cdot T_4) \big)\\
    \end{aligned}
    \end{equation}

  Because $\Phi_B^{m_2} \equiv \top$ and $(\eta_B = 0) \equiv \top$, we deduce:\\
  $\Phi_B^+(T_4) \equiv (\pii{B}{4} = 1) \wedge (C_{B,\{m_2\},0} > 0) \wedge (\pii{B}{4}' = \pii{B}{4} - C_{B,\{m_2\},0} \cdot T_4)$

  \item Computation of $\Rep_B$:
    \begin{equation}\label{}
      \Rep_B \equiv (\pii{A}{4} = \pii{A}{3}') \wedge (\pii{B}{4} =  1 - \pii{B}{3}')
    \end{equation}
    
  \item For this particular path, $\A(T_4) \equiv \top$.
\end{itemize}

We can deduce the value of the hybrid part $H_4$:

\begin{align}
\begin{split}
  H_4 &\equiv H_3 \wedge \big( \neg (C_{A,\{m_1,m_3\},2}<0) \vee \neg (\pii{A}{4}' < \pii{A}{4} - C_{A,\{m_1,m_3\},2} \cdot T_4) \big) \\
  &\quad \wedge (\pii{B}{4} = 1) \wedge (C_{B,\{m_2\},0} > 0) \wedge (\pii{B}{4}' = \pii{B}{4} - C_{B,\{m_2\},0} \cdot T_4) \\
  &\quad \wedge (\pii{A}{4} = \pii{A}{3}') \wedge (\pii{B}{3}' = 0)\\
\end{split} \\
\begin{split}
  H_4 &\equiv \Big( \neg (C_{B,\emptyset,0} > 0) \vee \neg (1 > \pii{B}{0}' - C_{B,\emptyset,0} \cdot T_1) \Big)\\
  &\quad \wedge (C_{A,\{m_1,m_3\},1} > 0) \wedge (\pii{A}{1}' = 1 - C_{A,\{m_1,m_3\},1} \cdot T_1) \wedge (\pii{A}{0}' = 0)\\
  &\quad \wedge \big( \neg (C_{A,\{m_1\},1}>0) \vee \neg (1 > \pii{A}{1}' - C_{A,\{m_1\},1} \cdot T_2) \big) \\
  &\quad \wedge \Big((C_{A,\emptyset,0} > 0) \vee \neg (C_{A,\{m_1\},1}<0) \vee \neg (1 < \pii{A}{1}' - C_{A,\{m_1\},1} \cdot T_2) \Big) \\
  &\quad \wedge (C_{B,\emptyset,1} < 0) \wedge (1 = 0 - C_{B,\emptyset,1} \cdot T_2) \\
  &\quad \wedge \big( \neg (C_{B,\{m_2\},1}<0) \vee \neg (0 < 1 - C_{B,\{m_2\},1} \cdot T_3) \big)\\
  &\quad \wedge \big( C_{A,\{m_1\},2} < 0 \wedge (\pii{A}{3}' = 0 - C_{A,\{m_1\},2} \cdot T_3) \big) \\
  &\quad \wedge \Big( \neg (C_{B,\{m_2\},1} > 0) \vee (0 > 1 - C_{B,\{m_2\},1} \cdot T_3) \Big)\\
  &\quad \wedge \big( \neg (C_{A,\{m_1,m_3\},2}<0) \vee \neg (\pii{A}{4}' < \pii{A}{3}' - C_{A,\{m_1,m_3\},2} \cdot T_4) \big) \\
  &\quad \wedge (C_{B,\{m_2\},0} > 0) \wedge (\pii{B}{4}' = 1 - C_{B,\{m_2\},0} \cdot T_4) \\
\end{split}
\end{align}

\subsection{Constraints for limit cycle}
\label{ap:limitCycle}

Here, we assumed that the path studied depicts the limit cycle.
Hence, the pre and postconditions apply to the same hybrid state $h_0 = h_4$,
which gives the following final set of constraints:
\begin{align}
  H_F &\equiv H_4 \wedge (\pii{A}{0} = \pii{A}{4}) \wedge (\pii{A}{0}' = \pii{A}{4}')
  \wedge (\pii{B}{0} = \pii{B}{4}) \wedge (\pii{B}{0}' = \pii{B}{4}') \\
\begin{split}
  H_F &\equiv \big( \neg (C_{B,\emptyset,0} > 0) \vee \neg (1 > \pii{B}{0}'
   - C_{B,\emptyset,0} \cdot T_1) \big)\\
  &\quad \wedge (C_{A,\{m_1,m_3\},1} > 0) \wedge (\pii{A}{1}' = 1 - 
  C_{A,\{m_1,m_3\},1} \cdot T_1) \\
  &\quad \wedge \big( \neg (C_{A,\{m_1\},1}>0) \vee \neg (1 > \pii{A}{1}' - 
  C_{A,\{m_1\},1} \cdot T_2) \big) \\
  &\quad \wedge \big((C_{A,\emptyset,0} > 0) \vee \neg (C_{A,\{m_1\},1}<0) \vee 
  \neg (1 < \pii{A}{1}' - C_{A,\{m_1\},1} \cdot T_2) \big) \\
  &\quad \wedge (C_{B,\emptyset,1} < 0) \wedge (1 = 0 - C_{B,\emptyset,1} \cdot 
  T_2) \\
  &\quad \wedge \big( \neg (C_{B,\{m_2\},1}<0) \vee \neg (0 < 1 - C_{B,\{m_2\},1} 
  \cdot T_3) \big)\\
  &\quad \wedge \big( C_{A,\{m_1\},2} < 0 \wedge (\pii{A}{3}' = 0 - C_{A,\{m_1\},2} 
  \cdot T_3) \big) \\
  &\quad \wedge \big( \neg (C_{B,\{m_2\},1} > 0) \vee (0 > 1 - C_{B,\{m_2\},1} 
  \cdot T_3) \big)\\
  &\quad \wedge \big( \neg (C_{A,\{m_1,m_3\},2}<0) \vee \neg (0 < \pii{A}{3}' - 
  C_{A,\{m_1,m_3\},2} \cdot T_4) \big) \\
  &\quad \wedge (C_{B,\{m_2\},0} > 0) \wedge (\pii{B}{0}' = 1 - C_{B,\{m_2\},0} 
  \cdot T_4) \enspace.
\end{split}
\end{align}